\documentclass[twocolumn,english]{article}
\usepackage[T1]{fontenc}
\usepackage[latin9]{inputenc}
\usepackage[a4paper]{geometry}
\usepackage{textcomp}
\usepackage{amsmath}
\usepackage{amssymb}
\usepackage{graphicx}

\makeatletter
\@ifundefined{date}{}{\date{}}
\makeatother

\usepackage{babel}
\begin{document}

\title{An Introduction to the Theory of Tachyons}

\author{Ricardo S. Vieira%
\thanks{E-mail: rsvieira@df.ufscar.br.%
} \vspace{0.3cm} \\  \textit{Departamento de Física, Universidade Federal de São Carlos} (UFSCar), \textit{São Carlos, Brasil}}
\maketitle
\begin{abstract}
The theory of relativity, which was proposed in the beginning of the
20th century, applies to particles and frames of reference whose velocity
is less than the velocity of light. In this paper we shall show how
this theory can be extended to particles and frames of reference which
move faster than light.
\end{abstract}

\section{The need for a theory of tachyons\label{sec: taquions }}

Scientists at the European Organization for Nuclear Research (CERN)
recently reported the results of an experiment \cite{CERN} on which
faster than light neutrinos were probably detected. Coincidentally,
about a week before the divulgation of this result, I was fortunate
to present at the 7th UFSCar Physics Week \cite{Vieira} a lecture
precisely on the subject of tachyons (the name given in theoretical
physics to faster than light particles). Students and Professors were
very interested on this lecture, specially after the news discussed
above. I have been encouraged since then to write a paper about the
ideas presented on that occasion, which constitute the present text.

Regardless of the results presented in \cite{CERN} are correct or
not, we point out that there are other experimental evidences for
existence of superluminal phenomena in nature \cite{bird}. Moreover,
the theory of tachyons could provide a better understanding of the
theory of relativity as well as some issues of quantum mechanics and
we believe that these arguments are enough for one to try to formulate
an \emph{extended theory of relativity}, which applies to faster than
light phenomena, particles and reference frames. Attempts to extend
the theory of relativity have been, of course, proposed by many scientists,
although the original sources are not easily accessible (in fact,
only recently I had knew about these works). Among the formulations
proposed, we highlight the works of Recami and collaborators \cite{Recami},
whose results coincide mostly with those who will be presented here.
Reference \cite{Recami} is a review on the subject, where the interested
reader will find a vast amount of references and may also get more
details about the theory, besides topics which will not be discussed
here (\emph{e.g.}, the tachyon electrodynamics).

For an extended theory of relativity, we mean a theory which applies
to particles and frames of reference moving with a velocity greater
than $c$ -- the speed of light in vacuum --, and also to particles
and reference frames ``moving'' back in time. In particular it is
necessary to extend the Lorentz transformations for such frames. Although
we have no problem at all to extend the Lorentz transformations in
a two-dimensional universe, $(x,t)$, we meet with certain difficulties
when we try to extend them in a universe of four dimensions, $(x,y,z,t)$.
The reasons why this happens will be commented in section \ref{sec: Lorentz4}.
Finally, in section \ref{sec: Lorentz6} we shall show that in a six-dimensional
world (3 space-like dimensions and other 3 time-like ones), we can
make those difficulties to disappear and therefore that becomes possible,
in this six-dimensional universe, to extend the Lorentz transformations
in agreement with the usual principles of relativity.

One of the reasons which makes the theory of tachyons almost unknown
is the (equivocated) belief that the theory of relativity forbids
the existence of faster than light particles, and that the light speed
represents an upper limit for the propagation of any phenomena in
nature. The argument commonly used to prove this statement is, as
stated by the theory of relativity, that no particle whatsoever can
be accelerated to reach or to exceed the speed of light, since this
would require spending an infinite amount of energy. This is not wrong,
but is also not completely correct. Indeed, this argument ignores
the possibility of these particles have been created at the same time
as the Big-Bang. In this way nobody had to speed them up -- they just
were born already with a greater than light speed.

Moreover, we can not rule out the possibility that such particles
might be created through some quantum process, analogous for instance
to the process of creation of particle-antiparticle pairs.

Finally, if we associate a complete isotropy and homogeneity to the
space-time, then it follows that none of their directions should be
privileged with respect to the others and thus the existence of faster
than light particles would naturally be expected instead of be regarded
as surprising. It is not the possibility of existence of tachyons
which requires explanation, on the contrary, an explanation must to
be given in the case of these particles \emph{do not }exist.

\section{On the space-time structure\label{sec: space-time}}

As is known, the formulation of the theory of relativity was due to
the efforts of several scientists (\emph{e.g.}, Lorentz, Poincaré,
Einstein, Minkowski etc.). The geometric description of the relativity
theory -- the so-called \emph{space-time theory} -- in its turn, was
first proposed by Poincaré \cite{Poincare} in 1905 and after independently
by Minkowski \cite{Minkowski} in 1909, where a more accessible and
detailed presentation of the subject was presented.

This geometric description, which contains the very essence of the
theory of relativity, may be grounded on the following statements,
or postulates%
\footnote{The influence of gravity will be explicitly neglected in this text.%
}:
\begin{enumerate}
\item The universe is a four-dimensional continuum -- three of them are
associated with the usual spatial dimensions \emph{X}, \emph{Y} and
\emph{Z}, while the other one it is associated with the time dimension;
\item The space-time is homogeneous and isotropic; 
\item The geometry of the universe is hyperbolic-circular. In the purely
spatial plans, \emph{XY}, \emph{YZ} and \emph{ZX} the geometry is
circular (\emph{i.e.}, euclidean), while in the plans involving the
time dimension,\emph{ }namely,\emph{ TX}, \emph{TY} and \emph{TZ},
the geometry is hyperbolic (\emph{i.e.}, pseudo-euclidean).
\end{enumerate}
In terms of the Poincaré-Minkowski description of space-time, any
inertial frame can be represented by an appropriate coordinate system,
which we shall call \emph{inertial coordinate system}. The movement
of a particle can be represented by a continuous curve -- a straight
line in the case of a free particle --, which we shall call the \emph{world-line}
of the particle. In particular, the velocity of a particle with respect
to a given frame of reference, say $R$, is determined by the direction
of the particle's world-line with respect to the time-axis of the
inertial coordinate system associated with $R$. Similarly, the relative
velocity $v$ between two frames $R$ and $R'$ is determined by the
direction of the time-axis of $R'$ with respect to that of $R$ and
a change of reference turns out to be, in this geometric description,
a mere hyperbolic rotation%
\footnote{In the case of $|v|>c$ we should consider a \emph{extended} hyperbolic
rotation as discussed in the section \ref{sec: Lorentz2}.%
} of the coordinate axes.

From these assumptions presented above, the theory of relativity can
be fully formulated. In particular, we point out that from these assumptions
we can deduce the principle of invariance of the speed of light (at
least in two dimensions). Indeed, the simple fact that the geometry
of space-time is hyperbolic implies the existence of a special value
of velocity which appears to be the same to all inertial frames, that
is, whose value does not change when one go from a inertial frame
to another. We can convince ourselves of this noting that in a hyperbolic
geometry there must exist certain lines (the asymptotes) which do
not change when a hyperbolic rotation is implemented. Therefore, if
there exist a particle whose world-line lies on those asymptotes,
the direction of the word-line of that particle should not change
by a hyperbolic rotation and, hence, its velocity must always be the
same for any inertial frame of reference. The experimental fact that
the speed of light is the same in any inertial frame provides, therefore,
a strong argument in favor of the hyperbolic nature of space-time.

For future use, we shall make some definitions and conventions which
will be used throughout the text. Since we intend to deal with particles
``moving'' in any direction of space-time, it is convenient to employ
a metric which is always real and non-negative. Let we define, therefore,
the metric by the expression 
\begin{equation}
ds=\sqrt{\left|c^{2}dt^{2}-dx^{2}-dy^{2}-dz^{2}\right|}.\label{eq:metrica}
\end{equation}
The choice of the metric, of course, does not affect the final results
of the theory, since we always have a certain freedom in defining
it.

In terms of the metric (\ref{eq:metrica}), we can classify the events
as time-like, light-like and space-like as the quantity $c^{2}dt^{2}-dx^{2}-dy^{2}-dz^{2}$
be positive, zero or negative, respectively. A similar classification
can be attributed to particles and frames of reference. Thus, for
instance, slower than light particles (bradyons) will be classified
as \emph{time-like} particles and faster than light particles (tachyons)
as\emph{ space-like} particles. Particles moving with the speed of
light (luxons) will be classified, of course, as \emph{light-like}
particles.

We can also classify the particles according to their ``direction
of movement'' in time. A particle which moves to the future will be
called a \emph{forward particle} and a particle moving to the past,
\emph{backward particle}. A particle with has an infinite velocity
exists only in the very present instant and we might call it a \emph{momentary}
\emph{particle}. A similar classification can be employed for reference
frames as well.

\section{The switching principle and antiparticles\label{sec: antiparticles}}

In the previous section we have introduced the concept of backward
particles as particles which travel back in time. Now we shall clarify
how we can interpret them from a physical point of view. However,
for the discussion becomes simpler we shall concern ourselves just
with time-like particles.

Let us begin by analyzing what must be the energy of a backward particle.
We know from the usual theory of relativity that the energy of a (time-like)
particle is related to its mass and its momentum through the expression
$E^{2}=p^{2}c^{2}+m^{2}c^{4}$. This quadratic expression for the
energy has two solutions: one of them represents the positive root
of that equation and the other the negative root (geometrically, this
equation describes, for a given $m$, the surface of a two-sheet hyperboloid).
In the theory of relativity we usually interpret the states of positive
energy as states which are accessible to any forward particle or,
which is the same, that a forward particle always has a positive energy.
Because of this association, we must for consistency regard negative
states of energy as accessible only to backward particles, so that
any backward particle has a negative energy.

These two separate concepts which not have a physical sense by themselves
-- namely, particles traveling backward in time and negative states
of energy (associated to free particles) -- can be reconciled through
what is called the \emph{switching principle}%
\footnote{Sometimes the terminology ``reinterpretation principle'' is employed%
}. This principle is based on the fact that any observer regard the
time as flowing from the past to the future and that any measurement
of the energy (associated to a free particle) results in a positive
quantity. Thus, the switching principle states that a backward particle
(whose energy is negative) should always be physically observed as
a typical forward particle (whose energy is positive).

\begin{figure}[h]
\begin{center}\includegraphics[scale=0.3]{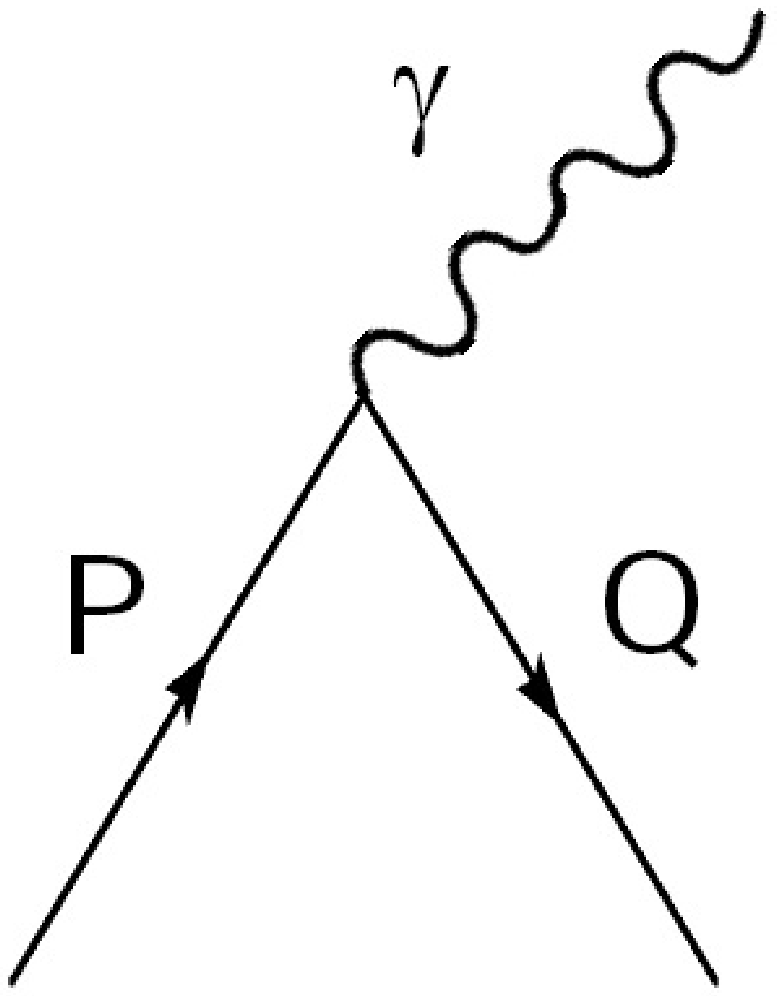}\end{center}

\caption{A forward particle $P$ interacts with a photon $\gamma$ and becomes
a backward particle $Q$.}
\label{figura: FD}
\end{figure}

Therefore, one might think that there are no differences at all between
forward and backward particles, since apparently the latter are always
seen in the same way as the first ones. However, this is not so because
whenever a backward particle is observed as a forward particle, some
of its properties turns out to be switched in the process of observation.
For instance, if a backward particle has a positive electric charge,
say $+e$, then, due to the conservation of electric charge principle,
we must actually observe a ``switched'' particle carrying the negative
charge $-e$. Let us clarify this point through the following experiment.

Consider the phenomenon described in the Figure \ref{figura: FD},
which describes a forward particle $P$ with electric charge $+e$
who interact at some time with a photon $\gamma$ and, by virtue of
this interaction, becomes a backward particle, $Q$. Note that the
particle $Q$ actually is the same particle $P$, but now it is traveling
back in time. Therefore, the actual electric charge of $Q$ is still
$+e$. Nevertheless, when this process is physically observed, the
observer should use (even unconsciously) the switching principle and
the phenomenon is to be interpreted as follows: two particles of equal
mass approach each to the other and, at some point, collide and annihilate
themselves, which gives rise to a photon. Since the photon has no
electric charge and the observed charge of the forward particle is
$+e$ it follows that the \emph{observed} charge of the backward particle
has to be $-e$. The conclusion which follows from this is that the
sign of the electric charge of a backward particle must be reversed
in the process of observation.

Thus, a backward particle of mass $m$ and electric charge $+e$ should
always be observed as a forward particle with the same mass and opposite
electric charge. But these properties are precisely the same as expected
to antiparticles. Therefore the switching principle enable us to interpret
a backward particle as an antiparticle. The concept of antiparticles
can be seen, hence, as a purely relativistic concept: it is not necessary
to talk about quantum mechanics to introduce the concept of antiparticles%
\footnote{The connection between backward particles and antiparticles was, of
course, proposed already by several scientists (\emph{e.g.}, Dirac
\cite{Dirac}, Stückelberg \cite{Stuckelberg,Stuckelberg-1}, Feynman
\cite{Feynman,Feynman-1}, Sudarshan \cite{Sudarshan}, Recami \cite{Recami}
etc.).%
}.

Finally, let us remark that these arguments are also valid in the
case of space-like particles, \emph{i.e.}, in the case of tachyons.
In the section \ref{sec: Dynamics} we shall see that the energy of
a tachyon is related to its momentum and mass through the relation
$E^{2}=p^{2}c^{2}-m^{2}c^{4}$, an equation which describes now a
single-sheet hyperboloid. From this we can see that tachyons must
have an interesting property: they can change its status of a forward
particle to the status of a backward one (and vice-versa) through
a simple continuous motion. In other words, by accelerating a tachyon
we can make it turn into an antitachyon and vice-versa (notice moreover
that at some moment the tachyon must becomes a momentary particle,
\emph{i.e.}, an particle with infinity velocity). This, of course,
it is only possible to space-like particles.

\section{Deduction of the extended Lorentz transformations (in two dimensions)\label{sec: Lorentz2}}

Let us show now how the Lorentz transformations can be generalized,
or extended, to frames of reference moving with a velocity greater
than that of light (as well as to frames of reference which travel
back in time). In this section we shall discuss however the theory
only in two dimensions. As commented before, we meet with several
difficulties to formulating a theory of tachyons in four dimensions
(these difficulties will be discussed in section \ref{sec: Lorentz4}).
We shall give here two deductions for the Extended Lorentz Transformations
(ELT), a algebraic deduction and a geometric one.

\textbf{Algebraic Deduction}: since in two dimensions the postulates
presented on the previous sections are sufficient to proof that light
propagates at the same speed $c$ for any inertial frame, we can take
this result as our starting point.

So, consider a certain event whose coordinates are $\left(ct,x\right)$
with respect to an inertial frame $R$ and $\left(ct',x'\right)$
with respect to another inertial frame $R'$ (which moves with the
velocity $v$ with respect to $R$). Suppose further that the coordinate
axes of these frames are always likely oriented and that the origin
of $R$ and $R'$ coincide when $t'=t=0$%
\footnote{Hereafter, whenever we speak in the frames of reference $R$ and $R'$
should be implicitly assumed that the relative velocity between them
is $v$ and that the conditions above are always satisfied.%
}.

Under these conditions, if a ray of light is emitted from the origin
of these frames at $t=0$, then this ray will propagates with respect
to $R$ according to the equation 
\begin{equation}
x^{2}-c^{2}t^{2}=0,\label{eq:luz}
\end{equation}
and for $R'$, by the principle of invariance of the speed of light,
also by 
\begin{equation}
x'^{2}-c^{2}t'^{2}=0.\label{eq:luz'}
\end{equation}
(\ref{eq:luz}) and (\ref{eq:luz'}) implies therefore, 
\begin{equation}
x'^{2}-c^{2}t'^{2}=\lambda(v)\left(x^{2}-c^{2}t^{2}\right),\label{eq:lambda1}
\end{equation}
where $\lambda(v)$ does not depend on the coordinates and the time,
but may depend on $v$.

In the other hand, since the frame $R$ moves with respect to $R'$
with the velocity $-v$, follows also that 
\begin{equation}
x^{2}-c^{2}t^{2}=\lambda(-v)\left(x'^{2}-c^{2}t'^{2}\right).\label{eq:lambda2}
\end{equation}
Therefore, from (\ref{eq:lambda1}) and (\ref{eq:lambda2}), we get
$\lambda(v)\lambda(-v)=1$. Besides, the hypothesis of homogeneity
and isotropy of space-time demands that $\lambda(v)$ does not depend
on the velocity direction%
\footnote{Indeed, only in this case the transformations do form a group, see
\cite{Poincare}. %
} and then we are led to the condition 
\begin{equation}
\lambda\left(v\right)^{2}=1\quad\Rightarrow\quad\lambda\left(v\right)=\pm1.
\end{equation}

Thus we have two cases to work on. Let us first analyze the first
case, namely, $\lambda\left(v\right)=+1$. In this case the equation
(\ref{eq:lambda1}) becomes 
\begin{equation}
x'^{2}-c^{2}t'^{2}=x^{2}-c^{2}t^{2},\label{eq:lambda+}
\end{equation}
whose solution, as it is known, is given by the usual Lorentz transformations,
\begin{equation}
ct'=\frac{ct-xv/c}{\sqrt{1-v^{2}/c^{2}}},\quad x'=\frac{x-vt}{\sqrt{1-v^{2}/c^{2}}}.\label{eq:Lorentz1}
\end{equation}
Note that these transformations contains the identity (for $v=0$)
and they are discontinuous only at $v=\pm c$. Consequently, these
transformations must be valid on the range $-v<c<v$ as well, but
nothing can be said for $v$ out from this range.

In this way, we hope that in the other case, \emph{i.e.}, when $\lambda\left(v\right)=-1$,
the respective transformations should be related to velocities greater
than that of light. Now we will show that this is indeed what happens.

To $\lambda\left(v\right)=-1$, equation (\ref{eq:lambda1}) becomes
\begin{equation}
x'^{2}-c^{2}t'^{2}=-\left(x^{2}-c^{2}t^{2}\right),\label{eq:lambda-}
\end{equation}
Through the formal substitutions $x\rightarrow\pm i\xi$ and $ct\rightarrow\pm ic\tau,$
we can rewrite (\ref{eq:lambda-}) as 
\begin{equation}
x'^{2}-c^{2}t'^{2}=\xi^{2}-c^{2}\tau^{2}.\label{eq:xi}
\end{equation}

\begin{figure}[h]
\begin{center}\includegraphics[scale=0.7]{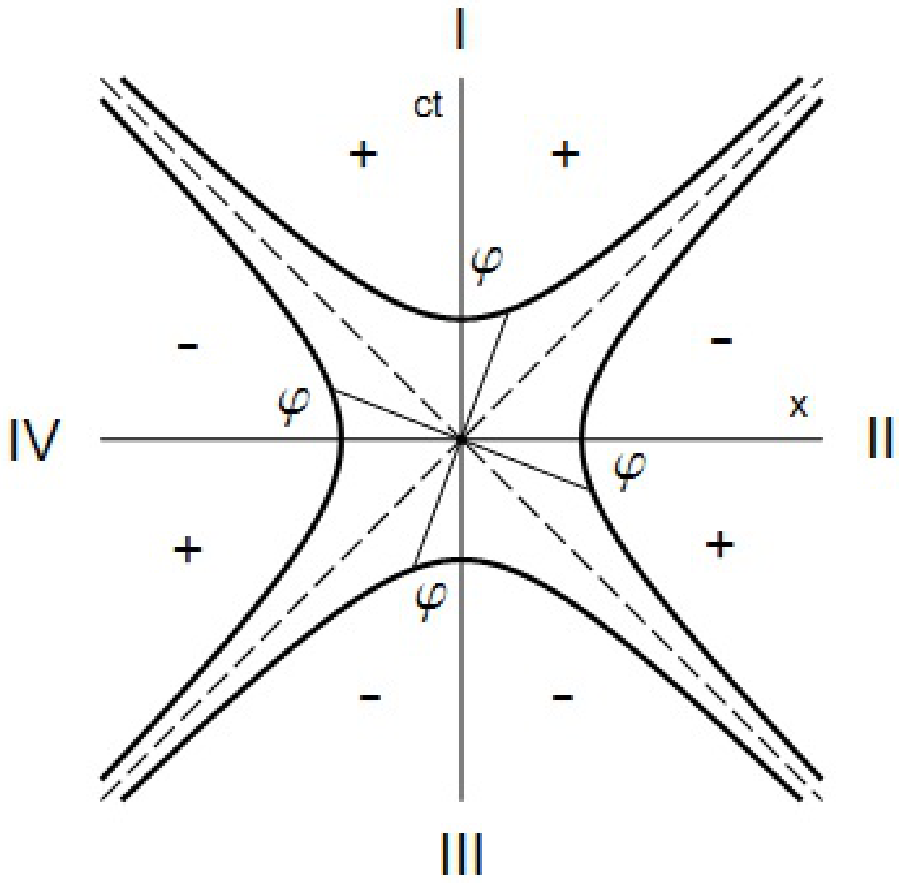}\end{center}\caption{The Curve $\left|c^{2}t^{2}-x^{2}\right|=\rho^{2}.$}
\label{figura: H}
\end{figure}

Equation (\ref{eq:xi}) has the same form as the equation (\ref{eq:lambda+})
and therefore has the same solution: 
\begin{equation}
ct'=\frac{c\tau-\xi v/c}{\sqrt{1-v^{2}/c^{2}}},\quad x'=\frac{\xi-v\tau}{\sqrt{1-v^{2}/c^{2}}}.
\end{equation}
Expressing them back in terms of $x$ and $t$, we get 
\begin{equation}
ct'=\pm\frac{ct-xv/c}{\sqrt{v^{2}/c^{2}-1}},\quad x'=\pm\frac{x-vt}{\sqrt{v^{2}/c^{2}-1}},\label{eq:TLE}
\end{equation}
and we have just to remove the signals ambiguity on the expressions
above. The correct sign, however, depends on the relatively direction
of the frames $R$ and $R'$ in their ``moviment'' on the space-time,
and can be seen in the Figure (\ref{figura: H}). In the case of a
\emph{forward} space-like transformation, it is easy to show that
the correct sign is the \emph{negative} one.

Note that these equations, as well as those of the previous case,
are discontinuous only for $v=\pm c$. But now they are real only
when $|v|>c$. The equations (\ref{eq:TLE}) represent thus the Lorentz
transformations between two space-like reference frames, that is,
the transformations associated to $|v|>c$, where the correct sign
depends if the frame $R'$ is a forward or backward reference frame
with respect to $R$ and can be determined by the Figure \ref{figura: H}.

\textbf{Geometrical Deduction}: from the geometrical point of view,
the ELT can be regarded as a (hyperbolic) rotation defined on the
curve%
\footnote{Such a rotation can be more elegantly described through the concept
of \emph{hyperbolic-numbers} \cite{h-numbers}. A hyperbolic number
is number of the form $z=a+hb$, where $\left\{ a,b\right\} \in\mathbb{R}$
and $h:\{h^{2}=+1,\: h\notin\mathbb{R}\}$. By defining the conjugate
$\bar{z}=a-hb$, it follows that $|\bar{zz}|=|a^{2}-b^{2}|=\rho^{2}$,
which represents an equation like (\ref{eq:Hip1}). This leads to
a complete analogy with complex numbers, but with the difference that
now these numbers describe a hyperbolic geometry. We also stress that
the same can be done through the elegant \emph{geometric algebra of
space-time} \cite{Vaz}, with the advantage that this formalism could
allow, perhaps, a generalization of these concepts to higher dimensions.%
} 
\begin{equation}
\left|c^{2}t^{2}-x^{2}\right|=\rho^{2}.\label{eq:Hip1}
\end{equation}
We can call such a transformation as an \emph{extended hyperbolic
rotation}. Note that (\ref{eq:Hip1}) represents a set of two orthogonal
and equilateral hyperbolæ. 
The asymptotes of this curve divide the plane onto four disjoint regions,
namely, the regions I, II, III and IV, as show in the Figure \ref{figura: H}.

To express such a rotation is convenient to introduce the \emph{extended
hyperbolic functions}, $\cosh\theta$ and $\sinh\theta$, defined
by the following relations 
\begin{equation}
ct=\rho\cosh\theta,\quad x=\rho\sinh\theta,
\end{equation}
where $\theta$ is the usual circular parameter so that $0\leq\theta<2\pi$
and $\rho$ is given by (\ref{eq:Hip1}). Note that in this geometric
description the velocity $v$ is given by 
\begin{equation}
v/c=\tanh\theta.\label{eq:vel}
\end{equation}

Expressions for $\cosh\theta$ and $\sinh\theta$ can be determined
in several ways. For instance, we can use the usual hyperbolic functions
$\cosh\varphi$ and $\sinh\varphi$ (where $\varphi$ is the usual
hyperbolic parameter so that $-\infty<\varphi<+\infty$) to define
them. Effectively, by introducing in each disjoint region of the space-time
a hyperbolic parameter $\varphi$, which must be measured as shown
in Figure \ref{figura: H}, we can see that the functions $\cosh\varphi$
and $\sinh\varphi$ can be used to parametrize each one of the four
branches of the curve (\ref{eq:Hip1}). Once specified the region
which $\theta$ belongs, $\rho$ and $\varphi$ determine in a unique
way any point of the curve (\ref{eq:Hip1}) and, therefore, they also
determine the extended hyperbolic functions. With these conventions,
we find out that 
\begin{equation}
\begin{array}{c}
\cosh\theta\equiv\begin{cases}
+\cosh\varphi, & \theta\in\text{I}\\
-\sinh\varphi, & \theta\in\text{II}\\
-\cosh\varphi, & \theta\in\text{III}\\
+\sinh\varphi, & \theta\in\text{IV}
\end{cases},\\
\\
\sinh\theta\equiv\begin{cases}
+\sinh\varphi, & \theta\in\text{I}\\
+\cosh\varphi, & \theta\in\text{II}\\
-\sinh\varphi, & \theta\in\text{III}\\
-\cosh\varphi, & \theta\in\text{IV}
\end{cases},
\end{array}\label{eq:FH1}
\end{equation}
where the parameters $\theta$ and $\varphi$ should be related by
the formula 
\begin{equation}
\tan\theta=\tanh\theta\equiv\begin{cases}
+\tanh\varphi, & \theta\in\left(\text{I, III}\right)\\
-\coth\varphi, & \theta\in\left(\text{II, IV}\right)
\end{cases}.\label{eq:thetaphi}
\end{equation}
Expressions for the extended hyperbolic functions can also be found
without making use of the usual hyperbolic functions. To do this,
we parametrize (\ref{eq:Hip1}) through the circular functions instead,
putting $ct=r\cos\theta$ and $x=r\sin\theta$, with $r=\sqrt{c^{2}t^{2}+x^{2}}$.
This allows us to write directly, 
\begin{equation}
\cosh\theta=\frac{\cos\theta}{\sqrt{\left|\cos2\theta\right|}},\quad\sinh\theta=\frac{\sin\theta}{\sqrt{\left|\cos2\theta\right|}}.\label{eq:FH2}
\end{equation}
or, in terms of the tangent,
\begin{equation}
\cosh\theta=\frac{\sigma(\theta)}{\sqrt{\left|1-\tan^{2}\theta\right|}},\quad\sinh\theta=\frac{\sigma(\theta)\tan\theta}{\sqrt{\left|1-\tan^{2}\theta\right|}},\label{eq:FH3}
\end{equation}
where 
\begin{equation}
\sigma(\theta)=\begin{cases}
+1, & -\pi/2<\theta<\pi/2\\
-1, & \pi/2<\theta<3\pi/2
\end{cases}.
\end{equation}

The equivalence between (\ref{eq:FH1}) and (\ref{eq:FH2}) or (\ref{eq:FH3})
is found when one takes into account (\ref{eq:thetaphi}).

Once defined the extended hyperbolic functions it is easy 
\begin{figure}[h]
\begin{center}\includegraphics[scale=0.5]{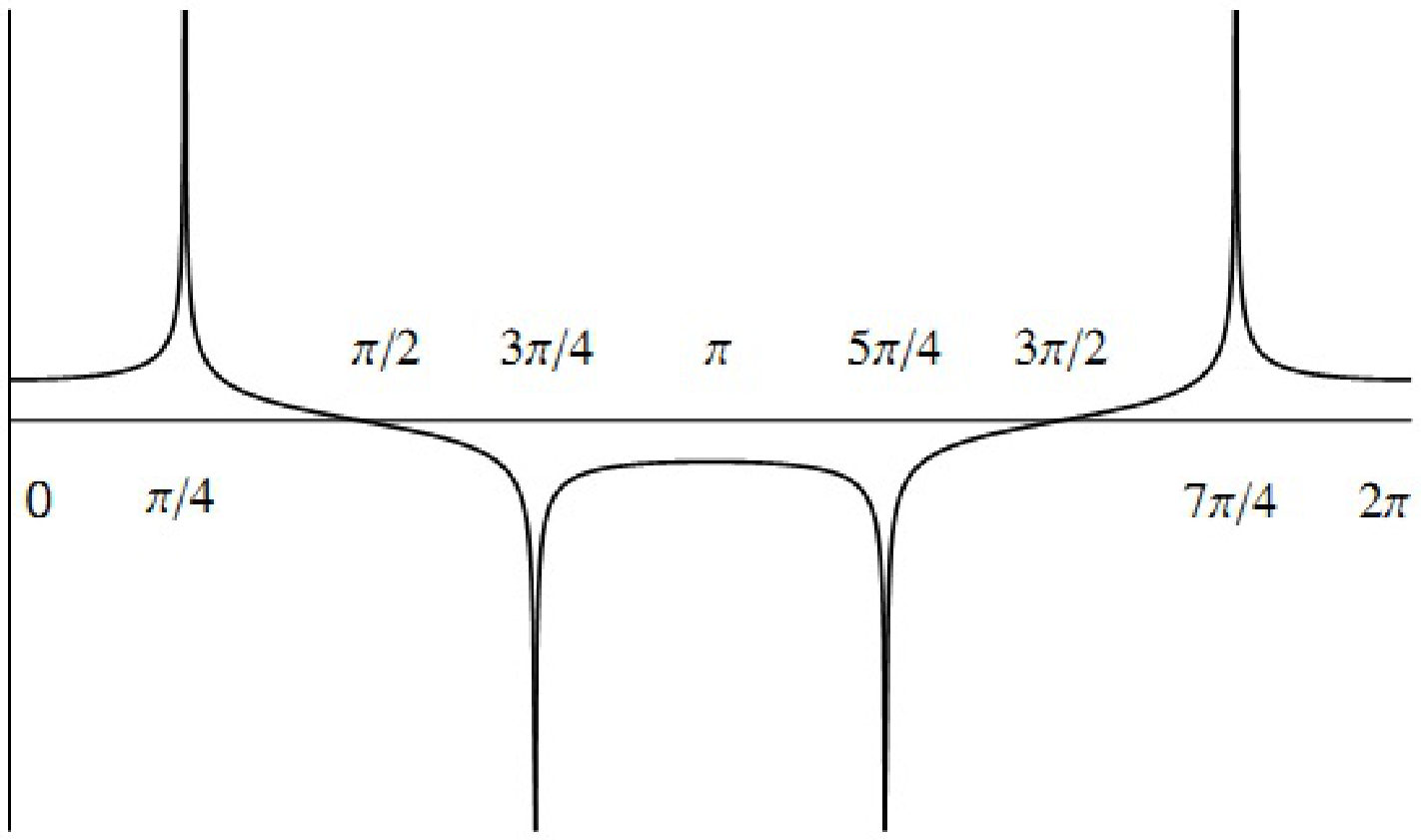}\end{center}\label{figura: C}

\caption{Graphic for the hyperbolic extended function $\text{\ensuremath{\cosh\theta}}$.
The graphic of $\sinh\theta$ is similar to this, but with a phase
difference of $\pi/2$ rad.}
\end{figure}
 to obtain expressions describing an extended hyperbolic rotation.
Let $\left(ct,x\right)=\left(\rho\cosh\theta_{1},\rho\sinh\theta_{1}\right)$
be the coordinates of a point on the plane with respect to a inertial
coordinate system $R$, which it is assumed to belongs to the sector
I of space-time. If we take a \emph{passive} hyperbolic rotation (\emph{i.e.},
if we rotate the coordinate axes), say by an angle $\theta_{12}$,
we shall obtain a new inertial coordinate system $R'$ and the coordinates
of that same point become $\left(ct',x'\right)=\left(\rho\cosh\theta_{2},\rho\sinh\theta_{2}\right)$
with respect to $R'$. Since $\theta_{12}=\theta_{1}-\theta_{2}$,
it follows that 
\begin{equation}
ct'=\rho\cosh\left(\theta_{1}-\theta_{12}\right),\quad x'=\rho\sinh\left(\theta_{1}-\theta_{12}\right).
\end{equation}

By replacing the expressions of $\cosh\left(\theta_{1}-\theta_{12}\right)$
and $\sinh\left(\theta_{1}-\theta_{12}\right)$ by any one of the
above expressions and by simplifying the resulting expressions, taking
also into account (\ref{eq:thetaphi}), we find out that $\left(ct',x'\right)$
is related to $\left(ct,x\right)$ through the equations 
\begin{equation}
\begin{array}{c}
ct'=\delta\left(\theta_{12}\right)\left(ct\cosh\theta_{12}-x\sinh\theta_{12}\right),\\
\\
x'=\delta\left(\theta_{12}\right)\left(x\cosh\theta_{12}-ct\sinh\theta_{12}\right),
\end{array}
\end{equation}
where 
\begin{equation}
\delta\left(\theta\right)=\begin{cases}
+1, & \tan^{2}\theta<1\\
-1, & \tan^{2}\theta>1
\end{cases}.\label{eq:delta}
\end{equation}
Finally, using (\ref{eq:FH3}) and putting $\tan\theta_{12}=v/c$
, we obtain directly the required transformations, which are identical
to those obtained before, namely, 
\begin{equation}
ct'=\varepsilon\left(\theta_{12}\right)\cfrac{ct-xv/c}{\sqrt{\left|1-v^{2}/c^{2}\right|}},\quad x'=\varepsilon\left(\theta_{12}\right)\cfrac{x-vt}{\sqrt{\left|1-v^{2}/c^{2}\right|}},\label{eq:TLE2}
\end{equation}
where we put $\varepsilon\left(\theta_{12}\right)=\sigma\left(\theta_{12}\right)\delta\left(\theta_{12}\right)$.
$\varepsilon\left(\theta_{12}\right)$ determines the correct sign
which must appear in front of these transformations, as can be visualized
in the Figure \ref{figura: H}.

\section{The composition law of velocities and inverse transformations\label{sec: velocity}}

The transformations deduced in the previous sections do form a group.
The ordinary Lorentz transformations is just a subgroup of it. Let
us demonstrate now this group structure.

First, note that the identity is obtained with $v=0$. We shall show
now that the composition of two ELT still results in another ELT.
For this we introduce a third inertial frame $R''$, which moves with
respect to $R'$ with velocity $u=c\tanh\theta_{23}$. $R'$ by its
turn is assumed to moves with the velocity $u=c\tanh\theta_{12}$
with respect to $R$. We already know the transformation law between
$R$ and $R'$, and we can write it down: 
\begin{equation}
\begin{array}{c}
ct'=\varepsilon\left(\theta_{12}\right)\cfrac{ct-x\tanh\theta_{12}}{\sqrt{\left|1-\tanh^{2}\theta_{12}\right|}},\\
\\
x'=\varepsilon\left(\theta_{12}\right)\cfrac{x-ct\tanh\theta_{12}}{\sqrt{\left|1-\tanh^{2}\theta_{12}\right|}}.
\end{array}\label{eq:C2}
\end{equation}
In its turn, the law of transformation from $R'$ to $R''$ is given
by an analogous expression: 
\begin{equation}
\begin{array}{c}
ct''=\varepsilon'\left(\theta_{23}\right)\cfrac{ct'-x'\tanh\theta_{23}}{\sqrt{\left|1-\tanh^{2}\theta_{23}\right|}},\\
\\
x''=\varepsilon'\left(\theta_{23}\right)\cfrac{x'-ct'\tanh\theta_{23}}{\sqrt{\left|1-\tanh^{2}\theta_{23}\right|}}.
\end{array}\label{eq:C1}
\end{equation}
In these equations $\varepsilon'\left(\theta_{23}\right)$ determines
the signals corresponding to the transformation from $R'$ to $R''$.
These signals, however, do not need to equal necessarily the signals
related to the transformation from $R$ to $R'$. Indeed, while in
the definition of $\varepsilon\left(\theta_{12}\right)$ the frame
of reference $R$ was supposed to belong to the region I of space-time,
the frame $R'$ might belong to any region of space-time. Thus, $\varepsilon'\left(\theta_{23}\right)$
should be regarded as a function to be yet determined.

Substitution of (\ref{eq:C2}) into (\ref{eq:C1}) gives us the law
of transformation between $R$ and $R''$. After some simplifications,
one can verify that the resulting expressions have the same form of
the ELT, namely 
\begin{equation}
\begin{array}{c}
ct''=\varepsilon''\left(\theta_{13}\right)\cfrac{ct-x\tanh\left(\theta_{13}\right)}{\sqrt{\left|1-\tanh^{2}\left(\theta_{13}\right)\right|}}\\
\\
x''=\varepsilon''\left(\theta_{13}\right)\cfrac{x-ct\tanh\left(\theta_{13}\right)}{\sqrt{\left|1-\tanh^{2}\left(\theta_{13}\right)\right|}}.
\end{array}\label{eq:C3}
\end{equation}
where 
\begin{equation}
\varepsilon''\left(\theta_{13}\right)=\frac{\varepsilon\left(\theta_{12}\right)\varepsilon'\left(\theta_{23}\right)}{\delta\left(\theta_{12},\theta_{23}\right)}
\end{equation}
with 
\begin{equation}
\delta\left(\theta_{12},\theta_{23}\right)=\begin{cases}
+1, & \tanh\theta_{12}\tanh\theta_{23}<1\\
-1, & \tanh\theta_{12}\tanh\theta_{23}>1
\end{cases},
\end{equation}
and 
\begin{equation}
\tanh\left(\theta_{13}\right)=\frac{\tanh\theta_{12}+\tanh\theta_{23}}{1+\tanh\theta_{12}\tanh\theta_{23}}=\tanh\left(\theta_{12}+\theta_{23}\right).\label{eq:comp1}
\end{equation}
From (\ref{eq:thetaphi}) we can see that (\ref{eq:comp1}) consists
of a generalization of the addition formula for the hyperbolic tangent,
which reveals its geometric meaning. In terms of the velocity $v$,
equation (\ref{eq:comp1}) can be rewritten as 
\begin{equation}
w=\frac{u+v}{1+uv/c^{2}}.\label{eq:comp2}
\end{equation}

Equation (\ref{eq:comp2}) expresses the composition law of velocities
in this extended theory of relativity. It is exactly the same as predicted
by the usual theory of relativity, but now it applies to any value
of velocity.

Let we show also that the inverse transformation does exist. To do
this, we impose onto (\ref{eq:C3}) the conditions $ct''=ct$ and
$x''=x$ and require that the resulting transformation be the identity.
For this it is necessary to have $\theta_{23}=-\theta_{12}$ and $\varepsilon''\left(\theta_{13}\right)=1$,
which enable us to evaluate $\varepsilon'\left(\theta_{23}\right)$
from the resulting expression of $\varepsilon''\left(\theta_{13}\right)$.
In fact, we find that 
\begin{equation}
\varepsilon'\left(\theta_{23}\right)=\delta\left(\theta_{12}\right)\big/\varepsilon\left(\theta_{12}\right)=\sigma\left(\theta_{12}\right),
\end{equation}
since $\delta\left(\theta_{12},-\theta_{12}\right)=\delta\left(\theta_{12}\right)$,
with $\delta\left(\theta_{12}\right)$ given (\ref{eq:delta}).

Substituting this result into (\ref{eq:C1}) we obtain the required
expressions for the inverse transformations, which when expressed
in terms of the velocity $v$ are given by 
\begin{equation}
\begin{array}{c}
ct=\varepsilon^{-1}\left(\theta_{12}\right)\cfrac{ct'+x'v/c}{\sqrt{\left|1-v^{2}/c^{2}\right|}},\\
\\
x=\varepsilon^{-1}\left(\theta_{12}\right)\cfrac{x'+vt'}{\sqrt{\left|1-v^{2}/c^{2}\right|}},
\end{array}\label{eq: inversas}
\end{equation}
and where we put $\varepsilon^{-1}\left(\theta_{12}\right)=\sigma\left(\theta_{12}\right)$. 

Note that the signs appearing on the inverse transformation are different
from that present on the direct transformations. This difference is
a consequence of what was commented before, that in the transformation
from $R$ to $R'$ we had assumed the starting frame $R$ always belonging
to the region I of space-time, while in the transformation from $R'$
to $R$ it is the frame $R'$ which is fixed on the region I. When
$|v|<c$ this asymmetry has no effect at all, since in this case the
signals are always the same in both expressions. But when $|v|>c$
however, we should alert that the inverse transformations can not
be simply obtained by replacing $v$ by $-v$. It is still necessary
multiply them by $-1$.

Finally, we mention that the associativity of ELT can be shown in
a similar way, which completes the group structure of the ELT.

\section{Conjugate Frames of Reference\label{sec: conjugated}}

We shall introduce now an important concept which can only be contemplated
in an extended theory of relativity: the concept of \emph{conjugate
frames of reference}. The definition is the following: two frames
of reference are said to be conjugate if their relative velocity is
infinite. Thus, if $v$ is the velocity of the frame $R'$ with respect
to the frame $R$, the conjugate frame of reference associated to
$R'$ is another reference frame $R^{*}$, whose velocity is $w=c^{2}/v$
when measured by $R$. In fact, we obtain from (\ref{eq:comp2}),
\begin{equation}
\lim_{u\rightarrow\infty}\left(\frac{u+v}{1+uv/c^{2}}\right)=\frac{c^{2}}{v}.
\end{equation}

Conjugate frames of reference are important because a space-like Lorentz
transformation between two frames, say, from $R$ to $R'$, can be
obtained also by a usual Lorentz transformation between $R$ and $R^{*}$.
For this, we simply have to replace $v\rightleftharpoons c^{2}/2$,
$ct^{*}\rightleftharpoons x$ and $x^{*}\rightleftharpoons ct$. In
fact, since $w=c^{2}/v$ is less than $c$ for $v>c$, it follows
that the transformation from $R$ to $R^{*}$ is given by 
\begin{equation}
ct^{*}=\frac{ct-xw/c}{\sqrt{1-w^{2}/c^{2}}},\quad x^{*}=\frac{x-wt}{\sqrt{1-w^{2}/c^{2}}}.
\end{equation}
Making the replacements indicated above we can see in this way that
we shall get the correct transformations between $R$ and $R'$.

From a geometrical point of view the passage of a given frame of reference
to its conjugate consists of a reflection of the coordinate axes relatively
to the asymptotes of the curve (\ref{eq:Hip1}), since this reflection
precisely has the the effect of changing $ct$ by $x$ and vice-versa
(and thus the effect of replacing $v$ by $c^{2}/v$). Therefore,
we can see that an extended Lorentz transformation can be reduced
to a usual Lorentz transformation by performing appropriate reflections
relatively to asymptotes (in case of a space-like transformation)
and around the origin (for a backward time-like transformation). This
gives us a new way to derive the ELT.

It is interesting to notice also that, if a particle has velocity
$u=c^{2}/v$ with respect to the frame $R$, then the velocity of
this particle for the frame $R'$ will be infinite. In other words
this particle becomes a momentary particle to the frame $R'$. More
important than that, if the particle velocity $u$ is greater than
$c^{2}/v$, and $v<c$, this particle becomes a backward particle
to $R'$ and it should be observed as an antiparticle by this reference
frame. On the other hand, if $v>c$ the frame of reference $R'$ should
observe an antiparticle whenever $u<c^{2}/v$.

In an analogous way, since the frame $R'$ moves with the velocity
$-v$ with respect to $R$ , it follows also that a particle with
velocity $u'=c^{2}/v$ must have an infinite velocity with respect
to $R$. So, in the case of $|v|<c$, the reference frame $R$ should
observe an antiparticle if $u'<c^{2}/v$ and, in the case of $|v|>c$,
only if $u'>-c^{2}/v$. These relationships might be, of course, more
easily found by analyzing (\ref{eq:comp1}) or (\ref{eq:comp2}).

\section{Rulers and clocks\label{sec: rulers}}

Consider two identical clocks, one of them fixed in the frame $R$
and the other fixed to the frame $R'$. Further, consider that these
clocks are synchronized on $t=t'=0$, where $R$ and $R'$ are in
the same position. We wish to compare the timing rate of these clocks,
as measured by one of those frames. For example, suppose we want to
compare the rhythm of these clocks when the time intervals are always
measured by $R$. For this, suppose that the clock fixed on $R'$
takes the time $\tau'$ to complete a full period of oscillation.
The time $T$ corresponding to this period of time, but now measured
by $R$, can be found through the inverse transformations (\ref{eq: inversas}).
Since this clock is at rest with respect to $R'$, we should put $x'=0$
on the first of the formulæ (\ref{eq: inversas}) and we shall get
\begin{equation}
T=\varepsilon^{-1}\left(\theta\right)\frac{\tau'}{\sqrt{\left|1-v^{2}/c^{2}\right|}}.\label{eq:tempo}
\end{equation}

Then, we can verify that a forward moving clock (with respect to $R$)
becomes slower in measuring time than an identical clock at rest,
when the speed of the clock is less than that of light (as it is well-known).
But for a faster than light clock we get that it continues to be slower
for $\left|v/c\right|<\sqrt{2}$ and it becomes faster when $\left|v/c\right|>\sqrt{2}$.
It is interesting to note that for $\left|v/c\right|=\sqrt{2}$ both
clocks go back to work at the same timing rate. Moreover, in the case
of a backward moving clock, we can see from the switching principle
that this clock should work in the counterclockwise, which is due
to the fact that a backward-clock should mark the time from the future
to the past.

Let us now verify what we got when the clocks are appreciated by the
reference frame $R'$. In this case we should use the direct transformations
and thus we shall get

\begin{equation}
T'=\varepsilon\left(\theta\right)\frac{\tau}{\sqrt{\left|1-v^{2}/c^{2}\right|}},\label{eq:tempo2}
\end{equation}
where $\tau$ is the time spent by the clock fixed at $R$ (which
is moving with speed $-v$ with respect to $R'$) for it to complete
a full oscillation. Now, the signal appearing on (\ref{eq:tempo2})
is determined according to the Figure \ref{figura: H} and the analysis
becomes more or less complicated. Of course, we have no problems at
all when $|v|<c$, then we shall analyze only the case where $|v|>c$.
Suppose first that the reference frame $R'$ is a forward frame with
respect to $R$. In this case we find that $\varepsilon\left(\theta\right)=-1$
and the moving clock will work in the opposite direction as compared
to the clock fixed at $R'$. This means that the clock at $R$ is
a backward-clock with respect to $R'$. We can convince ourselves
of this from what was commented in the previous section, where it
is necessary to put there $u=0$ and $|v|>c$ (and therefore $u<c^{2}/v$).

This is an interesting situation indeed, because we have just seen
that for the frame $R$, both clocks work clockwise (if the clocks
are forward ones) or counterclockwise (if the clocks are backward
ones) when $|v|>c$. In the other hand, for the frame $R'$ if its
own clock is working clockwise, then the moving clock should work
counterclockwise and vice-versa. In the case where the reference frame
$R'$ is backward with respect to $R$ these asymmetries persists
yet, but now is the frame $R$ which will see both clocks working
differently, while for $R'$ they will work accordingly. These asymmetries,
of course, just express the fact that the extended Lorentz transformations,
the direct and inverse one, are asymmetric by themselves in the case
of $|v|>c$.

Let we consider now two identical rulers, one of them placed at rest
in the frame $R$ and the other placed at rest with respect to $R'$.
We wish to compare the length of these rulers, when analyzed by one
of these frames. If $l_{0}'$ is the length of the ruler at $R'$,
when measured by this frame, the respective length $L$, as measured
by $R$, is obtained by determining where the extreme points of the
moving ruler is at a given instant $t$, say, $t=0$. Making use of
the second equation of the direct transformations , we find that 
\begin{equation}
L=\varepsilon\left(\theta\right)\cdot l_{0}'\sqrt{\left|1-v^{2}/c^{2}\right|},
\end{equation}

To $|v|<c$ we have the usual Lorentz contraction, but for $|v|>c$
we get that the moving ruler will be smaller than the ruler at rest
when $\left|v/c\right|<\sqrt{2}$ . The rulers go back once again
to have the same length when $\left|v/c\right|=\sqrt{2}$ and, finally,
for $\left|v/c\right|>\sqrt{2}$ they should present a ``Lorentz dilatation.''
Moreover, regarding $R'$ as a forward frame with respect to $R$,
it follows that for the frame $R$ the moving ruler will be oriented
contrarily with respect to the ruler at rest.

If, on the other hand, measurements are made by $R'$, then we find
that 
\begin{equation}
L'=\varepsilon^{-1}\left(\theta\right)\cdot l_{0}\sqrt{\left|1-v^{2}/c^{2}\right|},
\end{equation}
and now for the reference frame $R$ the ruler at motion (which has
the velocity $-v$) point out to the same direction as its ruler at
rest. We find again the same asymmetry commented above for the clocks.
These results can, of course, be more easily obtained -- and fully
understood -- through Minkowski diagrams.

\section{Dynamics\label{sec: Dynamics}}

In this section we intend to answer some questions concerning the
dynamics of tachyons. The expressions for the energy and momentum
for a faster than light particle will be deducted and we shall show
how these particles behave in the presence of a force field.

As a starting point we might assume that the principle of stationary
action also applies to faster than light particles. This, of course,
follows directly from the assumption of homogeneity and isotropy of
space-time, since we know that this principle is true for slower than
light for particles.

As one knows, the principle of stationary action states that there
exist a quantity $S$, called \emph{action}, which assumes an extreme
value (maximum or minimum) for any possible movement of a mechanical
system (in our case, a particle). On the other hand, in the absence
of forces, the motion of a particle corresponds to a space-time geodesic,
which reduces to a straight line by neglecting gravity. This means
that in case of a free particle the differential of action $dS$ should
be proportional to the line element $ds$ of the particle and we can
write in this way 
\begin{equation}
dS=\alpha ds,\quad ds=\sqrt{\left|c^{2}dt^{2}-dx^{2}-dy^{2}-dz^{2}\right|}.\label{eq:S}
\end{equation}
We must emphasize, however, that the constant of proportionality $\alpha$
can take different values at different regions of space-time, since
these regions are completely disconnected regions. Therefore, it is
convenient to consider each case separately.

In the case of a forward and time-like particle, (\ref{eq:S}) takes
the form 
\begin{equation}
dS=\mathcal{L}\left(u\right)dt,\quad\mathcal{L}\left(u\right)=\alpha c\sqrt{1-u^{2}/c^{2}},\label{eq:L1}
\end{equation}
where we had introduced the LaGrange function, $\mathcal{L}\left(u\right)$,
to express the action in terms of particle velocity.

As it is known, the expressions for the energy and momentum are obtained
by the formulas 
\begin{equation}
p\left(u\right)=\frac{\partial\mathcal{L}\left(u\right)}{\partial u},\quad E\left(u\right)=u\left[\frac{\partial\mathcal{L}\left(u\right)}{\partial u}\right]-\mathcal{L}\left(u\right).\label{eq:pE}
\end{equation}
Applying (\ref{eq:pE}) onto (\ref{eq:L1}) we obtain, thus 
\begin{equation}
p\left(u\right)=-\frac{\alpha u/c}{\sqrt{1-u^{2}/c^{2}}},\quad E\left(u\right)=-\frac{\alpha c}{\sqrt{1-u^{2}/c^{2}}}.
\end{equation}
To find $\alpha$ we may use the fact that for low speeds these expressions
should reduce to that obtained by Newtonian mechanics. Thus, for instance,
if we expand the expression for the momentum in a power series of
$u/c$ and we keep only the first term, we should get $p\approx-\alpha u/c$,
while Newtonian mechanics provides $p=mu$. Thus we get $\alpha=-mc$
and then 
\begin{equation}
p\left(u\right)=\frac{mu}{\sqrt{1-u^{2}/c^{2}}},\quad E\left(u\right)=\frac{mc^{2}}{\sqrt{1-u^{2}/c^{2}}},
\end{equation}
which are the same expressions of the usual theory of relativity.

In the case of a forward space-like particle (\emph{i.e}. in the case
of a forward tachyon), the LaGrange function takes the form 
\begin{equation}
\mathcal{L}\left(u\right)=\alpha c\sqrt{u^{2}/c^{2}-1}.
\end{equation}
and we get by (\ref{eq:pE}), the following expressions for momentum
and energy, 
\begin{equation}
p\left(u\right)=\frac{\alpha u/c}{\sqrt{u^{2}/c^{2}-1}},\quad E\left(u\right)=\frac{\alpha c}{\sqrt{u^{2}/c^{2}-1}}.
\end{equation}
The constant $\alpha$, however, no longer can be determined by comparing
these expressions with those obtained in Newtonian mechanics, since
the velocity of the particle is always greater than the speed of light
in this case. But we can stead evaluate the limit of these expressions
as $u\rightarrow\infty$, which give us 
\begin{equation}
\lim_{u\rightarrow\infty}p\left(u\right)=\alpha,\quad\lim_{u\rightarrow\infty}E\left(u\right)=0,
\end{equation}
by where we can see that $\alpha$ equals the momentum of a momentary
particle, that is, the momentum of a infinitely fast particle. 

\begin{figure}[h]
\begin{center}\includegraphics[scale=0.5]{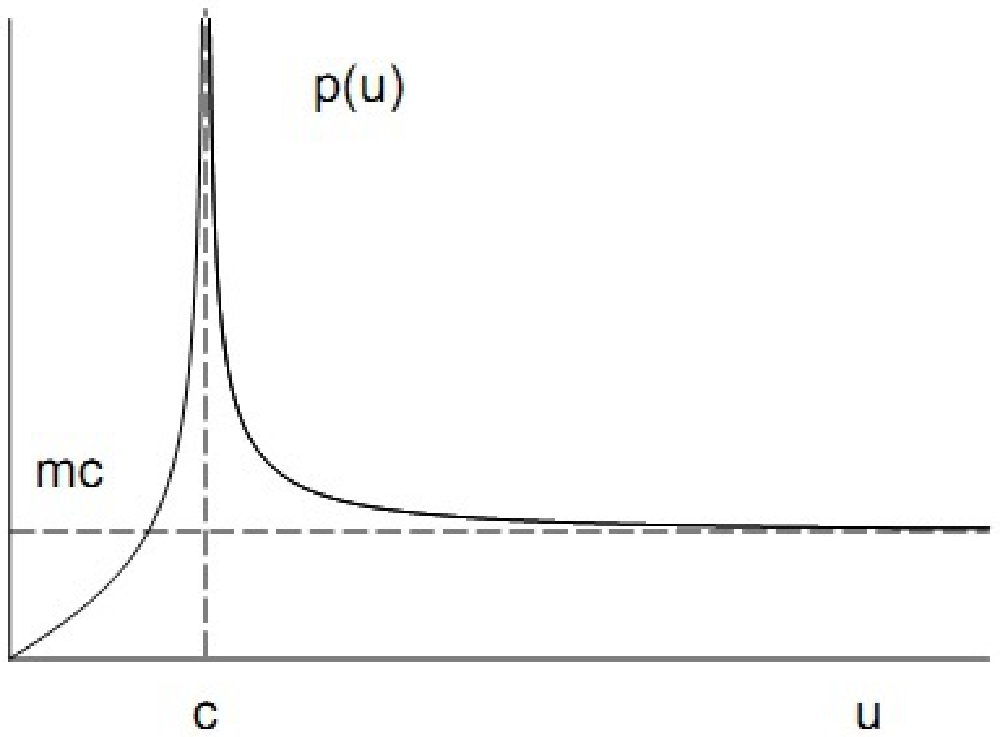} \includegraphics[scale=0.5]{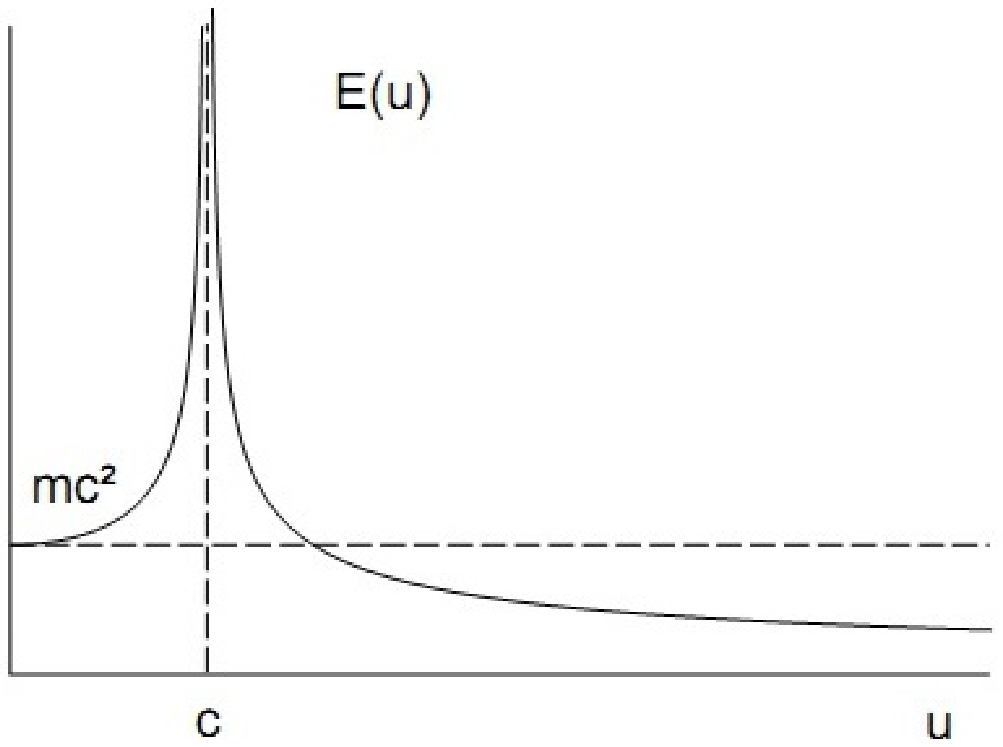}\end{center}\caption{Graphics for the momentum and energy of a forward particle with speed
$0\leq u<\infty$.}
\end{figure}

Since the mass of a particle must be a universal invariant, it follows
that we can also define a metric in the space of energy and momenta.
Making an analogy with $(\ref{eq:metrica})$, we can define this \emph{dynamic
metric }as 
\begin{equation}
c^{2}dm=\sqrt{\left|dE^{2}-c^{2}dp_{x}^{2}-c^{2}dp_{y}^{2}-c^{2}dp_{z}^{2}\right|}.
\end{equation}
Thus, it follows that the mass, the energy and the momentum must always
be related by the formula $mc^{2}=\sqrt{\left|E^{2}-c^{2}p^{2}\right|}$
. If we put $E=0$ in this last equation, we get $p=mc$ (we had assumed
that $p$ is positive for $u$ positive). We conclude in this way
that the momentum and energy of a tachyon should be given by the expressions
\begin{equation}
p\left(u\right)=\frac{mu}{\sqrt{u^{2}/c^{2}-1}},\quad E\left(u\right)=\frac{mc^{2}}{\sqrt{u^{2}/c^{2}-1}}.\label{eq:pE2}
\end{equation}

This result can also be obtained through the use of conjugate frames,
introduced in section \ref{sec: conjugated}. Suppose a particle with
velocity $u>c$ with respect to a reference frame $R$. Then, for
the conjugate reference frame $R^{*}$ this particle will have energy
$E^{*}=cp$ and momentum $p^{*}=E/c$, while its velocity becomes
$w=c^{2}/u$, which is less than $c$. But the expressions for momentum
and energy of a bradyon (\emph{i.e.}, a time-like particle) are given
by (\ref{eq:pE}) and we get, therefore 
\begin{equation}
p^{*}=\frac{mw}{\sqrt{1-w^{2}/c^{2}}},\quad E^{*}=\frac{mc^{2}}{\sqrt{1-w^{2}/c^{2}}}.
\end{equation}
We can verify in this way that the replacements of $E^{*}$, $p^{*}$
and $w$ into the above formulæ result exactly in the expression (\ref{eq:pE2}).

In the case of backward particles, the LaGrange function changes sign,
since $dt$ is negative and $\mathcal{L}=dS/dt$. Therefore, backward
particles must have a negative energy, a result which we had already
indicated in section \ref{sec: antiparticles}.

Finally, let us show how a space-like particle should behave when
subjected to a force field. The expression for the force acting on
the particle remains, of course, being given by Newton's Law $\dot{p}=F$,
but now $p$ is given by (\ref{eq:pE2}). Calculating the derivative,
we obtain 
\begin{equation}
F=-\frac{ma}{\left(u^{2}/c^{2}-1\right)^{3/2}},
\end{equation}
where the $a=du/dt$ denotes the acceleration of the particle. Note
that the force and the acceleration point out to opposite directions.
Consequently, two space-like particles that attract (repel) must go
far (go near) one from the other. The concepts of attraction/approximation
and repulsion/expulsion, therefore, are no longer equivalent concepts
when we deal with tachyons. In fact, if we try to approach (move away)
from a tachyon, its relative speed decrease (increase), an effect
that is contrary to our common sense but that can be demonstrated
by a simple analysis of (\ref{eq:comp2}).

\section{Causality and the Tolman Paradox\label{sec: Tolman}}

One of the arguments usually employed to show that faster than light
particles can not exist is that their existence would imply a violation
of the principle of causality. Indeed, suppose a tachyon is emitted
by a body $A$ at time $t_{A}$ and it is absorbed by another body
$B$ at time $t_{B}$, where $t_{B}>t_{A}$. Since these two events
(emission of a tachyon by $A$ and its absorption by $B$) are separated
by a space-like distance we know from the theory of relativity that
it is possible to find a determined frame of reference (whose speed
is less than that of light) where the chronological order of events
is reversed. Thus, if we assume the event $A$ as the cause of the
event $B$, we shall conclude then, for this moving frame of reference,
that the effect precedes its cause.

Therefore, if we admit the concepts of cause and effect as absolute
concepts and that the cause always precedes the effect, we would then
effectively be led to the conclusion that faster than light particles
can not exist. However, there is nothing from the mathematical point
of view that supports this hypothesis. On the contrary, if we assume
a complete isotropy and homogeneity of space-time, we have no choice
but to consider the concepts of cause and effect as relative concepts.

We already had seen, of course, other quantities which were regarded
as absolute in the usual theory of relativity and now they had to
be seen as relative quantities. An example of this is the concept
of emission and absorption. Indeed, since by the switching principle
we did not observe backward-particles but only forward-antiparticles,
the processes of emission (absorption) of backward-particles should
always be observed as an absorption (emission) of forward-antiparticles.
We can, of course, find out if a particle was actually emitted (absorbed)
by analyzing the process in the frame where the source (emitter) is
at rest, in this case we shall speak of an \emph{intrinsic} or \emph{proper}
emission (absorption).

Nevertheless, the relative nature of causality does not lead to any
contradiction into the theory. In nature, a phenomenon never describes
an isolated event but only a continuous succession of events, which
we shall call a\emph{ physical process}. Geometrically, a physical
process describes a continuous curve in space-time and therefore has
an absolute character: the order in time in which events occur may
differ from a reference frame to another but the curve itself, which
corresponds to the phenomenon in question, should be the same for
any of them. This is enough for instance to show that we can not go
back in time and kill our grandfather. In fact, the mere fact of our
existence means that there exists a curve in space-time connecting
our grandfather to us, and since this curve has an absolute character,
it can not be disconnected to any observer, even backing in time.

\begin{figure}[h]
\begin{center}

\includegraphics[scale=0.35]{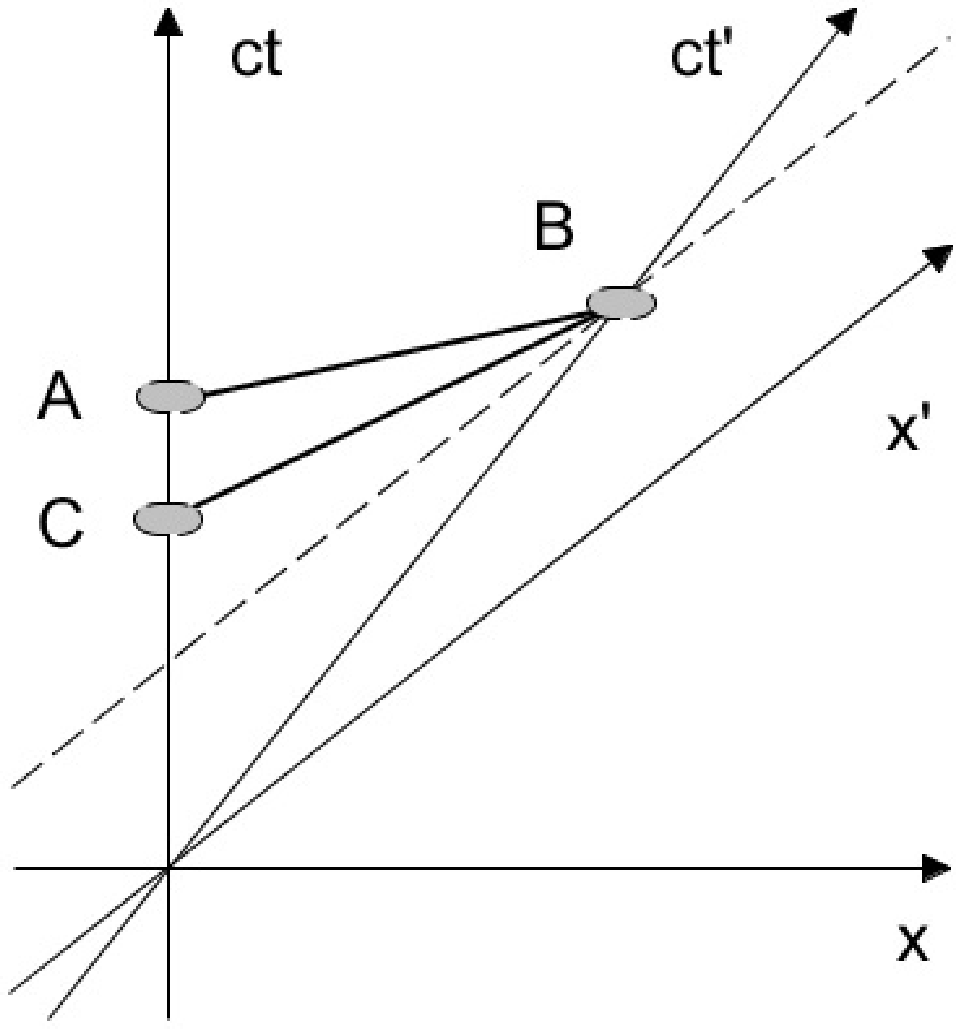} \hspace{0.2cm}\includegraphics[scale=0.35]{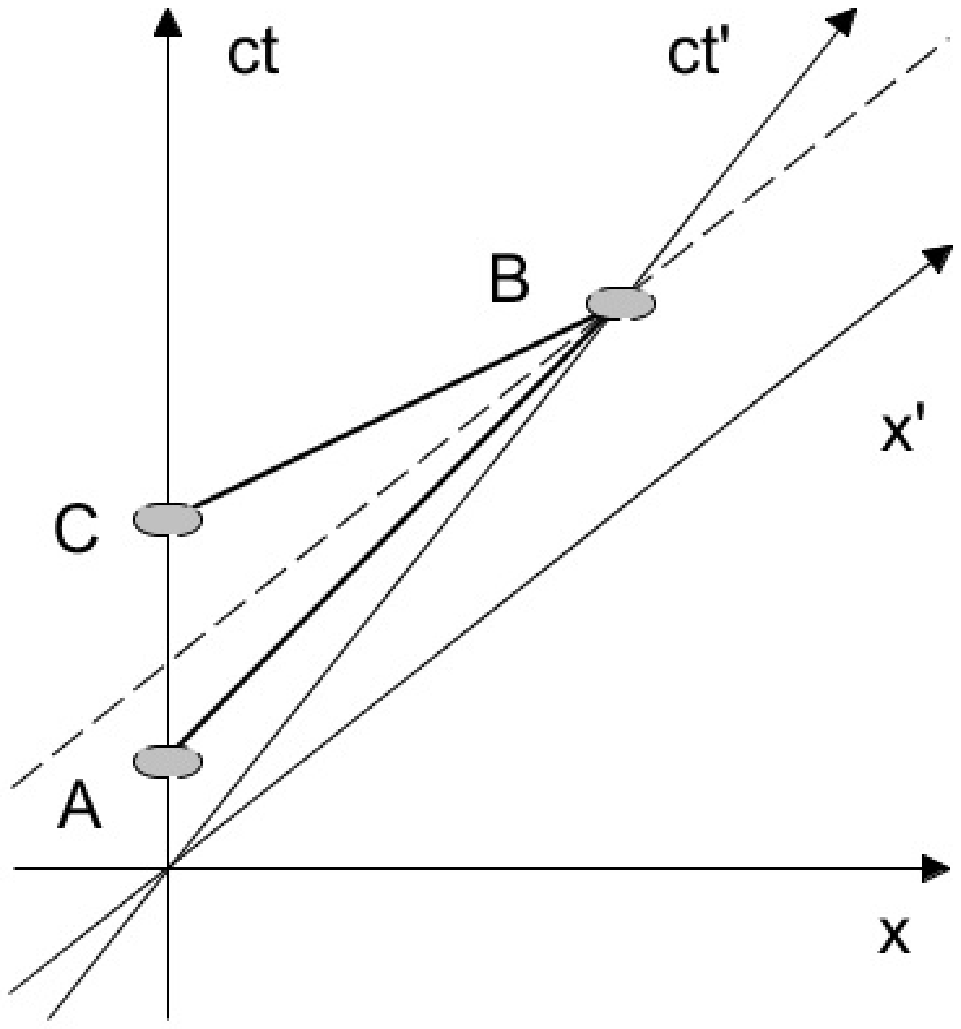}

\end{center}

\caption{Experiment devised by Tolman.}

\label{figura: T}
\end{figure}

As an example of a paradox which involves the concept of causality,
let us examine the interesting paradox proposed and discussed by Tolman
\cite{Tolman} in 1917 (although it had been pointed out ten years
earlier by Einstein \cite{Einstein}). Further details can be checked
in \cite{Bohm,Recami-1,Recami-2}.

The paradox consists in showing that if faster than light particles
does exist, then it would be possible to send information to the past.
To show this, consider the references $R$ and $R'$ (with $R'$ moving
with speed $v<c$ with respect to $R$) are equipped with some special
phone devices, whose communication is done through the transmission
of tachyons. Imagine, then, that an observer $R$ sends a message
(event $A$), for instance in the form of a question, to $R'$ and
that after he had received (event B), $R'$ sends back his answer
to $R$, which is then received by $R$ (event $C$). It turns out
that, as seen in Figure \ref{figura: T}a, the observer $R'$ can
calibrate its device in such a way that the message sent by him arrives
at $R$ even before $R$ had sent its question. Thus, since $R$ got
the answer before he had make the ask, he might decides to not do
the question, in which case we shall meet to a very paradoxical situation.

The failure on this argument is that it mixes the descriptions of
the reference frames when the terms ``emition'' and ``absorption''
are employed. In fact, we shall show now (following \cite{Recami})
that if the frame of reference $R'$ \emph{intrinsically} receives
the question submitted by $R$, as well as he \emph{intrinsically}
sends its answer, then the frame of reference $R$ should always receive
the reply after he had make the ask.

For this, suppose that the question of $R$ is sent with velocity
$u_{1}$ and the answer comes back with the velocity $u_{2}$, where
$u_{1}$ and $u{}_{2}$ are greater than the speed of light and they
are always measured by $R$. Notice then that for the answer of $R'$
arrives before the question sent by $R$ it is necessary to have $\left|u_{1}\right|>\left|u_{2}\right|$
because for $R$ the event $C$ occurs before $A$ and both arrive
on $B$ at the same time. Nevertheless, for the frame of reference
$R'$ to receive intrinsically the question sent by $R$ it is necessary
to have $\left|u_{1}\right|<c^{2}/v$, since in the other case, as
we have seen in section \ref{sec: conjugated}, this message would
be a backward-message to $R'$ and hence $R'$ would simply see the
emission of the message by transmission of antitachyons as predicted
by the switching principle. Similarly, for $R'$ intrinsically to
send its answer, we must have $\left|u_{2}\right|>c^{2}/v$, since
the answer is sent in the negative direction of the axis $X$ of $R$.
Thus, we had show that $\left|u_{2}\right|>\left|u_{1}\right|$, a
contradiction.

Therefore, if the reference $R'$ effectively receives the question
sent from $R$ and effectively sends her answer, the process must
be described by the Figure \ref{figura: T}b, not by the Figure \ref{figura: T}a.
In this case, the answer always comes after the question sent by $R$.

Of course, the Figure \ref{figura: T}a also represents a valid physical
process, but not one as described by the Tolman paradox. Each frame
of reference should have its own view of what is actually happening
in it. For instance, for $R$ the experiment described by Figure \ref{figura: T}a
corresponds to the case where there is, first, the emission of the
message $C$ (through an antitachyon jet) and soon after, there is
the emission of the message $A$ (via a tachyon jet); finally, both
messages come together at $B$. But for $R'$ all happens as he had
sent the message $C$ to $R$ by an antitachyon jet and, at the same
time, sent the message $A$ by means of a tachyons jet; then after
some time the observer $R$ would receive the message $C$ and then
the message $A$. Note that the chronological order of the events
depends on the frame considered, but the process itself (which connects
the events $A$, $B$ and $C$) is unique.

\section{Difficulties for formulating the theory of tachyons in four dimensions\label{sec: Lorentz4}}

In the section \ref{sec: Lorentz2} we show that in two dimensions
the Lorentz transformations can be extended in order to relate any
imaginable pair of inertial frames. When, however, we try to do the
same in four dimensions we meet with serious difficulties, which we
shall briefly discuss now.

Let us remind the reader that by extending the Lorentz transformations
in two dimensions, we had made use of an important principle, which
states that the speed of light is the same in any inertial frame.
We saw, moreover, that this principle can be deduced from the postulates
presented in section \ref{sec: space-time}, and hence this principle
can be seen as a theorem of the relativity theory. When, however,
we pass to a universe in four dimensions, is no longer possible to
proof the general validity of this theorem, in the same way as we
did in two dimensions. Actually, in four dimensions we can not even
say that light travels in spherical surfaces with speed $c$ in all
inertial frame of reference (it is only in the case of $v<c$ we can
mathematically proof this assertion, from the postulates of relativity).

To show more clearly why we meet with these difficulties, let us consider
for example a source of light fixed on the origin of the frame $R$
and suppose that with respect to this frame light propagates in spherical
surfaces with velocity $c$. In four-dimensional space-time the asymptotes
are replaced by a cone -- the light cone. The geometric region of
space-time associated to a beam of light (emitted by the source fixed
at the origin of the reference frame $R$) always belongs to this
light cone.

The usual theory of relativity also shows that for any other time-like
frame of reference, the propagation of light is still represented
by the same light cone. But how should the light behave itself with
respect to a space-like frame, say, $R'$? In this case, the time
axis of this frame always point out to the outside region of the light
cone of $R$, so that is not evident that the light emitted by the
source fixed at $R$ should also propagates in spherical surfaces
with speed $c$ with respect to $R'$. Beside this, we might ask what
happens when a light source is fixed on $R'$. Does the light emitted
by this source propagates in spherical surfaces with speed $c$ with
respect to $R'$? And what is to be said with respect to the frame
$R$, where the source has velocity $v>c$, does the light propagates
spherical waves with speed $c$?

As one can see, there are several issues that can not be answered
immediately, not without having additional information. Indeed, the
behavior of light depends on its intrinsic properties, and these properties
are determined by a extended theory of electromagnetism which is \emph{a
priori} unknown. Let us examine these possibilities a little more.

We can suppose for instance that the source of light fixed at $R'$
(which we shall suppose to have a velocity greater than $c$, with
respect to $R$) also emits spherical waves with speed $c$, when
measured by $R'$. So, there must exist now another light cone corresponding
to space-like frames of reference, a cone which must contour the axis
of time of $R'$. Since the axis of time of $R$ always should be
outside from this new light cone, it follows that for $R$ the light
emitted by the moving source will not propagates into spherical surfaces
of speed $c$, but rather it will take a form of a two-sheet hyperboloid
surface. In this case the speed of light would depend of the direction
and space-time could no longer be considered isotropic anymore.

If, in other hand, we regard that the light emitted by the source
fixed on $R'$ also propagates in spherical surfaces with respect
to the reference frame $R$, that is, that there are a unique light
cone in the universe by where the light always ``travels'', no matter
what is the speed of the source, then we can see that, although for
$R$ the light propagates in spheres of speed $c$, now the same is
not true for $R'$, which should observes the light to propagating
in a two-sheet hyperboloid.

For light spread in spherical surfaces and with velocity $c$ to both
reference frames is necessary that the transverse coordinates, $y'$
and $z'$ of the frame $R'$ be imaginary quantities (assuming $y$
and $z$ real). Indeed, just in this case we can map a ``time-like
light cone'' to a ``space-like light cone.'' Effectively, this can
be shown by deducing the ELT in four dimensions, assuming from the
start the validity of the principle of invariance of the speed of
light to any inertial frame. We have in this case, 
\begin{equation}
x'^{2}+y'^{2}+z'^{2}-c^{2}t'^{2}=\lambda\left(v\right)\left(x^{2}+y^{2}+z^{2}-c^{2}t^{2}\right),
\end{equation}
in place of (\ref{eq:lambda1}). As it is known, the solution of this
equation for $\lambda\left(v\right)=+1$ consists of the usual Lorentz
transformations, while for $\lambda\left(v\right)=-1$ we shall get
\begin{equation}
\begin{array}{c}
ct'=\pm\cfrac{ct-xv/c}{\sqrt{v^{2}/c^{2}-1}},\quad x'=\pm\cfrac{x-vt}{\sqrt{v^{2}/c^{2}-1}},\\
\\
y'=\pm iy,\quad z'=\pm iz,
\end{array}\label{eq:TLE4D}
\end{equation}
and the transverse coordinates $y'$ and $z'$ become imaginary. The
introduction of imaginary coordinates is, however, deprived of physical
sense%
\footnote{Some authors, for instance \cite{Dawe}, argue that the introduction
of imaginary coordinates in the theory does not constitute a problem.
The solution proposed by them consists by regarding the coordinate
axis $Y'$ and $Z'$ themselves as imaginary axes, so that the observers
in the (space-like) frame of reference $R'$ always should interpret
the coordinates $y'$ and $z'$ as \emph{relatively} real quantities.
This, however, is not consistent. Indeed, let we take for instance
a ray of light which propagates on the $Y$ direction with respect
ot $R$. For the frame of reference $R'$, the light travels through
an oblique trajectory, and the components of its velocity will have
the magnitudes $\left|c_{x}'\right|=v$ and $\left|c_{y}'\right|=c\sqrt{v^{2}/c^{2}-1}$.
Thus, for the ray of light to propagate with speed $c'=c$, it is
necessary that the square of $c_{y}'$ be a negative number, which
contradicts the hypothesis that $c_{y}'$ is a \emph{relatively} real
number .%
}.

If, by its turn, we deny the validity of the principle of invariance
of the speed of light, then we obtain the transformations 
\begin{equation}
\begin{array}{c}
ct'=\pm\cfrac{ct-xv/c}{\sqrt{v^{2}/c^{2}-1}},\quad x'=\pm\cfrac{x-vt}{\sqrt{v^{2}/c^{2}-1}},\\
\\
y'=\pm y,\quad z'=\pm z.
\end{array}
\end{equation}
where now all coordinates are real. The problem here is that we lie
in the cases discussed above, where the space-time can not be considered
isotropic.

Note that in each one of the possibilities discussed above, the transverse
components of the physical quantities become different. Thus, for
example, the formulation of the electrodynamics of tachyons will assume
different forms in each one of these formulations. Only the experiment
can decides which one or another is the correct, (assuming that someone
of them are correct).

In short, the arguments presented above show us that in a four-dimensional
universe we can not extend the Lorentz transformations in order to
satisfy all the postulates presented in section \ref{sec: space-time},
unless it is introduced imaginary quantities, which has no physical
interpretation. We shall see in the next section that, if we not impose
the validity of the postulate $1$, \emph{i.e}. that the universe
is four-dimensional, then it becomes possible to build up a theory
of tachyons which satisfy the other postulates fully.

\section{An possible theory in six dimensions\label{sec: Lorentz6}}

We might wonder why we find difficulties on extending the Lorentz
transformations in a four-dimensional universe if in a two-dimensional
universe this generalization is straightforward. We can figure out
what is happening by reflecting a little about. The reason by which
this occurs lies in the fact that in four dimensions the number of
space-like dimensions is different from the number of time-like dimensions
(3 space-like dimensions against 1 time-like dimension). Since experience
shows us that the universe has three space-like dimensions (at least),
it follows that we must consider a six-dimensional universe (at least)
-- 3 space-like dimensions against other 3 time-like dimensions%
\footnote{The conception of a six-dimensional universe also was already proposed
before, see \cite{Recami}.%
}. 

\begin{figure}[h]
\begin{center}\includegraphics[scale=0.4]{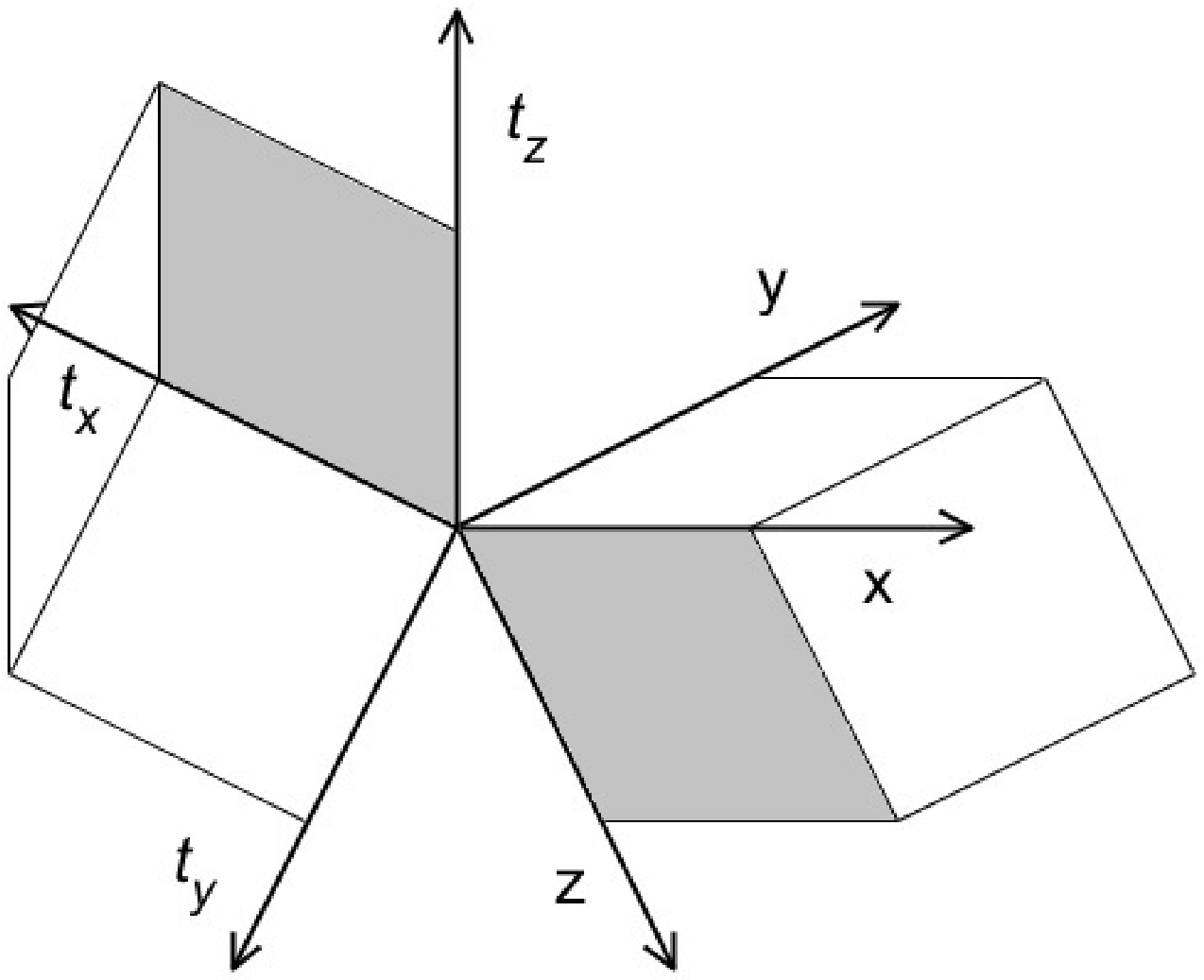}\end{center}

\caption{An illustration for the six-dimensional universe -- ``orthogonal universes.''}

\label{figura: 6}
\end{figure}

But what is to be understood by a universe with three time-like dimensions?
The interpretation we propose here is the following: although this
universe has three time-like dimensions, the \emph{physical time},
that is, the time which is actually measured by an observer, is always
a one-dimensional quantity. Indeed, this physical time must match
with the proper-time of the observer, that is, it should be determined
by the world line length of the observer concerned.

Thus, for consistence, we should consider also that the other two
time-like dimensions which are orthogonal to the observer's world
line, should be always \emph{inaccessible} to this observer. Therefore,
although we regard the universe as six-dimensional in nature, the
\emph{physical universe} happens to be just four-dimensional. In such
a way, this six-dimensional universe can be interpreted as consisting
of two orthogonal dimensional universes of three dimension each%
\footnote{The concept of ``paralel universes'' has been proposed, of couse,
several times by the scientic fiction stories, but I cannot remember
any of them who had introduced the concept of ``orthogonal universes.'' %
}.

These interpretations enable us to define a six-dimensional metric
for this universe by the expression 
\begin{equation}
ds=\sqrt{\left|c^{2}dt_{x}^{2}+c^{2}dt_{y}^{2}+c^{2}dt_{z}^{2}-dx^{2}-dy^{2}-dz^{2}\right|},
\end{equation}
and a \emph{physical metric} by 
\begin{equation}
d\sigma=\sqrt{\left|c^{2}dt_{x}^{2}-dx^{2}-dy^{2}-dz^{2}\right|},\label{eq:MO}
\end{equation}
where we suppose $t_{x}$ as the physical time measured by the observer
(by a convenient choice of the coordinate system, this can always
be done, of course).

Let us then show now how one can get the ELT in complete agreement
with the relativity postulates presented in section \ref{sec: space-time}.
For this, let a particular event has coordinates $\left(ct_{x},ct_{y},ct_{z},x,y,z\right)$
with respect to $R$, but suppose furthermore that only the coordinates
$\left(ct_{x},x,y,z\right)$ are accessible to this reference frame.
Similarly, for the reference frame $R'$ let $\left(ct_{x}',ct_{y}',ct_{z}',x',y',z'\right)$
be the coordinates of that same event, and suppose now that only the
coordinates $\left(ct_{x}',x',y',z'\right)$ are accessible to $R'$.

Let we analyze first the case where $v<c$. It is evident in this
case that the accessible dimensions to $R$ and $R'$ should coincide
themselves (in fact, this is true for $v=0$ and therefore the result
must continuity requirement). Thus, we obtain directly that the transformations
should be just the usual Lorentz transformations, plus the relations
\begin{equation}
ct_{y}'=ct_{y},\quad ct_{z}'=ct_{z},\quad y'=y,\quad z'=z.
\end{equation}

Consider now the more interesting case where $v>c$. Here, on the
contrary, there must be a reversal of the accessible dimensions of
$R$ when they are observed by $R'$. Indeed, one might say, for instance,
that the frames of reference $R$ and $R'$ are now in differents
``orthogonal universes.'' Thus, the coordinates $ct_{y}$ and $ct_{z}$
should become accessible to $R'$, while the coordinates $y$ and
$z$ should become inaccessible to him. Note also that the necessary
condition for the light propagates in spherical surfaces with speed
$c$ in both frames of reference is only that one has $d\sigma'=d\sigma=0$
for rays of light. Nevertheless, we can consider a stronger hypothesis,
namely, that in the six-dimensional universe the light propagates
accordingly the equation $ds'=ds=0$. In this case it is easy to verify
that the previous condition is satisfied as well if we write 
\begin{equation}
ct_{y}'=\pm y,\quad ct_{z}'=\pm z,\quad y'=\pm ct_{y},\quad z'=\pm ct_{z},\label{eq:tytz}
\end{equation}
which must to be added to the expressions 
\begin{equation}
ct_{x}'=\pm\frac{ct_{x}-xv/c}{\sqrt{v^{2}/c^{2}-1}},\quad x'=\pm\frac{x-vt_{x}}{\sqrt{v^{2}/c^{2}-1}},\label{eq:TL4}
\end{equation}
already deduced in the section \ref{sec: Lorentz2}. These equations
constitute the ELT related to this six-dimensional universe. Notice,
in special, that all coordinates are real.

Let us finally interpret the results. In the first place, is easy
to see that in both frames of reference light propagates with speed
$c$. This can be done by direct substitution of (\ref{eq:tytz})
and (\ref{eq:TL4}) into (\ref{eq:MO}). In addition, notice that
in this formulation, tachyons might be observed in the reference $R$
with a shape very different from what they present to $R'$ because
to the reference $R$ the transverse coordinates are given by $y=\pm ct_{y}'$
and $z=\pm ct_{z}'$, which are not accessible to $R'$.

Further, it is interesting to notice that if a source of light is
fixed to the frame $R'$, then the frame $R$ should see the light
propagating in spherical surfaces with speed $c$, as we had shown.
But since the source has in this case a velocity greater than $c$,
we must have the formation of a Mach cone in fact, since the source
is always ahead from the waves which it emits. One can show that the
superposition of waves emitted by a superluminal source results at
two front waves whose shapes have the form of a two-sheet hyperboloid
surface. The group velocity of these waves depends on the direction,
being always larger than $c$ (except in the direction $X$ whose
velocity equals $c$). This, however, is due purely to an interference
phenomenon, unlike the cases of the previous formulations, which this
phenomenon was due to a anisotropy of space-time itself. 

This kind of waves are often called ``$X$ waves'', they are superluminal
solutions of the Maxwell's equations \cite{Rodrigues,Recami-3} --
these waves have indeed been observed and even produced in the laboratory
\cite{bird,Rodrigues,Recami-3,CH1}. This is an important experimental
verification of the existence of superluminal phenomena in nature.

\section*{Acknowledgments}

The Author is grateful to Prof. Dr. A. Lima-Santos by reading the
manuscript and by their suggestions. The Author also thanks to the
Fundação de Amparo à Pesquisa do Estado de São Paulo for financial
support.

\end{document}


\title{Uma Introdução à Teoria dos Táquions\\\foreignlanguage{english}{{\large{}
(An Introduction to the Theory of Tachyons)}}}

\author{Ricardo S. Vieira%
\thanks{E-mail: rsvieira@df.ufscar.br.%
} \vspace{0.3cm} \\  \textit{Departamento de Física, Universidade Federal de São Carlos, São Carlos, SP, Brasil}}
\maketitle
\begin{abstract}
A teoria da relatividade, proposta no início do século XX, aplica-se
à partículas e referenciais cuja velocidade é menor que a velocidade
da luz. Nesse artigo mostraremos como esta teoria pode ser estendida
à partículas e referenciais que se movimentam mais rapidamente que
a luz. \vspace{0,3cm}\\ \foreignlanguage{english}{The theory of
relativity, which was proposed in the beginning of the 20th century,
applies to particles and frames of reference whose velocity is less
than the velocity of light. In this paper we shall show how this theory
can be extended to particles and frames of reference which move faster
than light.}
\end{abstract}

\section{Sobre a necessidade de se formular uma teoria dos táquions\label{sec: taquions }}

Cientistas do Centro Europeu de Pesquisas Nucleares (CERN) divulgaram
recentemente os resultados de um experimento \cite{CERN} onde provavelmente
foram detectados neutrinos \emph{mais rápidos que a luz}. Por coincidência,
cerca de uma semana antes da divulgação desse resultado, tive a felicidade
de apresentar na 7ª Semana da Física da UFSCar \cite{Vieira} uma
palestra justamente sobre o tema dos táquions (nome dado em física
teórica às partículas mais rápidas que a luz). Essa palestra gerou
grande interesse por parte de professores e alunos, o qual foi ainda
mais aumentado após a divulgação da notícia comentada acima. Desde
então tenho sido incentivado a escrever um artigo sobre as ideias
apresentadas naquela ocasião, o que me propus a fazer nessas últimas
semanas e cujo resultado constitui o presente texto. 

Estando ou não corretos os resultados apresentados em \cite{CERN},
mencionamos que existem outras evidências experimentais sobre a existência
de fenômenos superluminais na natureza \cite{bird} e, independentemente
disso, que a teoria dos táquions permite uma melhor compreensão da
própria teoria da relatividade, bem como de alguns temas de mecânica
quântica. Acreditamos que esses sejam motivos suficientes para se
formular uma \emph{teoria da relatividade estendida} que se aplique
também a fenômenos, partículas e referenciais mais rápidos que a luz.
Tentativas de se estender a teoria da relatividade já foram, é claro,
propostas por diversos cientistas, muito embora as fontes originais
não sejam de fácil acesso (com efeito, só tive conhecimento desses
trabalhos recentemente). Dentre as formulações propostas, destacamos
a formulação apresentada por Recami \cite{Recami} e colaboradores,
cujos resultados coincidem em sua grande maioria com os que serão
apresentados aqui. A referência \cite{Recami} consiste em um artigo
de revisão sobre o assunto, onde o leitor interessado encontrará uma
vasta quantidade de referências e poderá obter também mais detalhes
sobre a teoria, além de tópicos que não serão discutidos aqui (\emph{e.g.},
a eletrodinâmica de táquions). 

Por uma teoria da relatividade estendida, queremos nos referir a uma
teoria que se aplique à partículas e referenciais que se movimentam
com velocidades maiores que a da luz, bem como à partículas e referenciais
que se ``movimentam'' para trás no tempo. Em especial é necessário
estender as transformações de Lorentz para tais referenciais. Embora
essa extensão possa ser deduzida, sem maiores problemas, em um universo
de duas dimensões, $(x,t)$, nos deparamos com dificuldades ao tentarmos
estendê-las em um universo de quatro dimensões, $(x,y,z,t)$. As razões
pelas quais isso ocorre serão comentadas na seção \ref{sec: Lorentz4}.
Por fim, na seção \ref{sec: Lorentz6} mostraremos que em um universo
de seis dimensões (com três dimensões espaciais e três temporais),
àquelas dificuldades desaparecem e torna-se possível estender as transformações
de Lorentz de acordo com os princípios da relatividade usual. 

Um dos motivos que tornam a teoria dos táquions quase que desconhecida
é a convicção (equivocada) de que a teoria da relatividade proíbe
a existência de partículas mais velozes que a luz, e que essa velocidade
representa um limite máximo para a propagação de qualquer fenômeno.
Um dos argumentos geralmente empregados para provar essa afirmação
é que, conforme estabelece a teoria da relatividade, nenhuma partícula
pode ser acelerada até atingir, ou ultrapassar, a velocidade da luz,
já que para isso seria necessário se gastar uma quantidade infinita
de energia. Isso não está errado, porém, também não está \emph{completamente
correto}. Com efeito, esse argumento não leva em conta a possibilidade
de que essas partículas possam ter sido criadas no exato momento da
criação do universo -- assim, ninguém teve de acelerá-las, elas simplesmente
já nasceram com uma velocidade maior que a da luz. 

Além disso, não podemos descartar a possibilidade de que tais partículas
possam ser criadas através de algum processo quântico, análogo por
exemplo ao processo de criação de pares de partículas-antipartículas. 

Por fim, se associarmos uma completa isotropia e homogeneidade ao
espaço-tempo, então segue que nenhuma de suas direções deve ser privilegiada
em relação às outras e, portanto, a existência de partículas mais
rápidas que a luz deveria ser naturalmente esperada, ao invés de ser
encarada com surpresa. Não é a possibilidade de existência dos táquions
que requer explicação, pelo contrário, uma explicação deveria ser
dada caso essas partículas não existissem.

\section{Sobre a estrutura do espaço-tempo\label{sec: space-time}}

Como se sabe, a formulação da teoria da relatividade deveu-se ao esforço
de vários cientistas (\emph{e.g}., Lorentz, Poincaré, Einstein, Minkowski
etc.). A descrição geométrica da teoria da relatividade -- a chamada
teoria do espaço-tempo --, por sua vez, foi proposta primeiramente
por Poincaré \cite{Poincare} em 1905 e, depois, de maneira mais acessível
e detalhada por Minkowski \cite{Minkowski} em 1909. 

Essa descrição geométrica, que contém a própria essência da teoria
da relatividade, pode ser fundamentada nas seguintes afirmações, ou
postulados%
\footnote{A influência da gravidade será explicitamente desprezada nesse texto.%
}.
\begin{enumerate}
\item O universo é um continuum em 4 dimensões -- três dessas dimensões
estão associadas às dimensões espaciais usuais $X$, $Y$ e $Z$,
enquanto que a outra está associada a uma dimensão temporal, $T$. 
\item O espaço-tempo é homogêneo e isotrópico.
\item A geometria do universo é hiperbólico-circular. Nos planos puramente
espaciais, $XY$, $YZ$ e $ZX$ a geometria é circular (\emph{i.e.},
euclidiana), enquanto que nos planos que envolvem a dimensão temporal,
$TX$, $TY$ e $TZ$, a geometria é hiperbólica (\emph{i.e.}, pseudo-euclidiana). 
\end{enumerate}
Em termos da descrição de Poincaré-Minkowski, qualquer referencial
inercial pode ser representado por um sistema de coordenadas adequado,
que chamaremos de sistema de coordenadas inercial. O movimento de
uma partícula fica representada por uma curva contínua -- uma reta,
caso a velocidade da partícula seja constante --, a que chamamos de
\emph{linha de universo} da partícula. Em especial, a velocidade de
uma partícula determina-se pela tangente do ângulo (hiperbólico) que
sua linha de universo faz com o eixo do tempo do sistema de coordenadas
no qual a velocidade é referida. Analogamente, a velocidade relativa
entre dois referenciais, $v$, é determinada pela tangente do ângulo
(hiperbólico) entre os eixos de tempo dos sistemas de coordenadas
correspondentes e uma mudança de referencial consiste, nessa descrição
geométrica, a uma mera rotação hiperbólica%
\footnote{No caso em que $\left|v\right|>c$ devemos considerar uma rotação
hiperbólica \emph{estendida}, conforme será discutido na seção \ref{sec: Lorentz2}.%
} dos eixos coordenados.

Através dos postulados apresentados acima, toda a teoria da relatividade
pode ser formulada. Em especial, destacamos que a partir desses postulados
podemos mostrar o princípio de invariância da velocidade da luz (pelo
menos em duas dimensões). De fato, o simples fato de que a geometria
do espaço-tempo é hiperbólica implica a existência de uma classe de
partículas cuja velocidade é sempre a mesma para qualquer referencial
inercial. Podemos nos convencer disso ao notar que em uma geometria
hiperbólica existem certas retas (as assíntotas) que não se alteram
quando uma rotação hiperbólica é implementada. Se, portanto, a linha
de universo de uma partícula for paralela a uma dessas assíntotas,
a sua direção será sempre a mesma para qualquer sistema de coordenadas
inercial e, consequentemente, a sua velocidade também será sempre
a mesma para qualquer referencial inercial. O fato experimental de
que da velocidade luz é a mesma em qualquer referencial inercial fornece,
assim, um forte argumento a favor do caráter hiperbólico do espaço-tempo.

Para uso futuro, faremos algumas definições e convenções que serão
utilizadas durante o texto. Uma vez que estamos dispostos a levar
em consideração partículas que se ``movimentam'' em qualquer direção
do espaço-tempo, convém empregarmos uma métrica que seja sempre real
e não-negativa. Optamos, assim, por definir a\emph{ métrica} pela
expressão,
\begin{equation}
ds=\sqrt{\left|c^{2}dt^{2}-dx^{2}-dy^{2}-dz^{2}\right|}.\label{eq:metrica}
\end{equation}
A escolha da métrica, é claro, não interfere nos resultados finais
da teoria, uma vez que se tem certa liberdade em defini-la. 

Em termos da métrica (\ref{eq:metrica}), vamos classificar os eventos
como do tipo tempo, luz ou espaço conforme a quantidade $c^{2}dt^{2}-dx^{2}-dy^{2}-dz^{2}$
seja positiva, nula ou negativa, respectivamente. Uma classificação
semelhante pode ser atribuída a partículas e referenciais. Assim,
por exemplo, partículas mais lentas que a luz (brádions) serão classificadas
como partículas do \emph{tipo tempo} e partículas mais rápidas que
a luz (táquions) como partículas do\emph{ tipo espaço}. Partículas
que se movem com a velocidade da luz (luxons) serão classificadas,
é claro, como\emph{ }partículas do \emph{tipo luz}.

Também classificaremos as partículas de acordo com a sua ``direção
de movimento'' no tempo. Partículas que se movimentam para o futuro
serão chamadas de \emph{progressivas} e as que se movimentam para
o passado, de \emph{retrógradas}. Partículas que se movimentam com
velocidade infinita só existem no instante presente e por isso serão
chamadas de \emph{momentâneas}. Uma classificação idêntica pode ser
empregada à referenciais.

\section{Antipartículas e o Princípio da Inversão\label{sec: antiparticles}}

Na seção anterior introduzimos o conceito de partículas retrógradas
como partículas que voltam no tempo. Nessa seção vamos esclarecer
como elas podem ser interpretadas do ponto de vista físico. Para que
a discussão se torne mais simples, entretanto, vamos considerar apenas
partículas do tipo tempo, isto é, partículas cuja velocidade é menor
que a da luz.

Deixe-nos começar o nosso estudo analisando qual deve ser a energia
de uma partícula retrógrada. Sabemos da teoria da relatividade usual
que a energia de uma partícula (do tipo tempo) relaciona-se com sua
massa e seu momentum através da expressão $E^{2}=p^{2}c^{2}+m^{2}c^{4}$.
Essa expressão quadrática tem duas soluções para a energia: uma raiz
positiva e outra negativa (geometricamente, esta equação descreve,
para $m$ fixo, um hiperboloide de duas faces). Na teoria relatividade
geralmente interpretamos os estados de energia positiva como estados
acessíveis a qualquer partícula progressiva, ou, em outras palavras,
que uma partícula progressiva possui sempre uma energia positiva.
Em virtude dessa associação, decorre que, por consistência, devemos
interpretar os estados de energia negativa como acessíveis somente
às partículas retrógradas, ou, em outros termos, que qualquer partícula
retrógrada deve possuir uma energia negativa.

Esses dois conceitos que separadamente não possuem sentido físico
-- a saber, partículas que voltam no tempo e partículas livres de
energia negativa --, podem ser conciliados através do que se chama
de \emph{princípio da inversão}%
\footnote{\emph{Switching principle}, em inglês. Às vezes o termo \emph{``princípio
da reinterpretação''} é também empregado.%
}. Esse princípio baseia-se no fato de que qualquer observador considera
o tempo como que fluindo do passado para o futuro e que qualquer medição
da energia de uma partícula livre resulta em uma quantidade positiva.
Desse modo, o princípio da inversão estabelece que uma partícula retrógrada
(cuja energia é negativa) deve sempre ser fisicamente observada como
uma partícula progressiva usual (cuja energia é positiva).

\begin{figure}[h]
\begin{center}\includegraphics[scale=0.3]{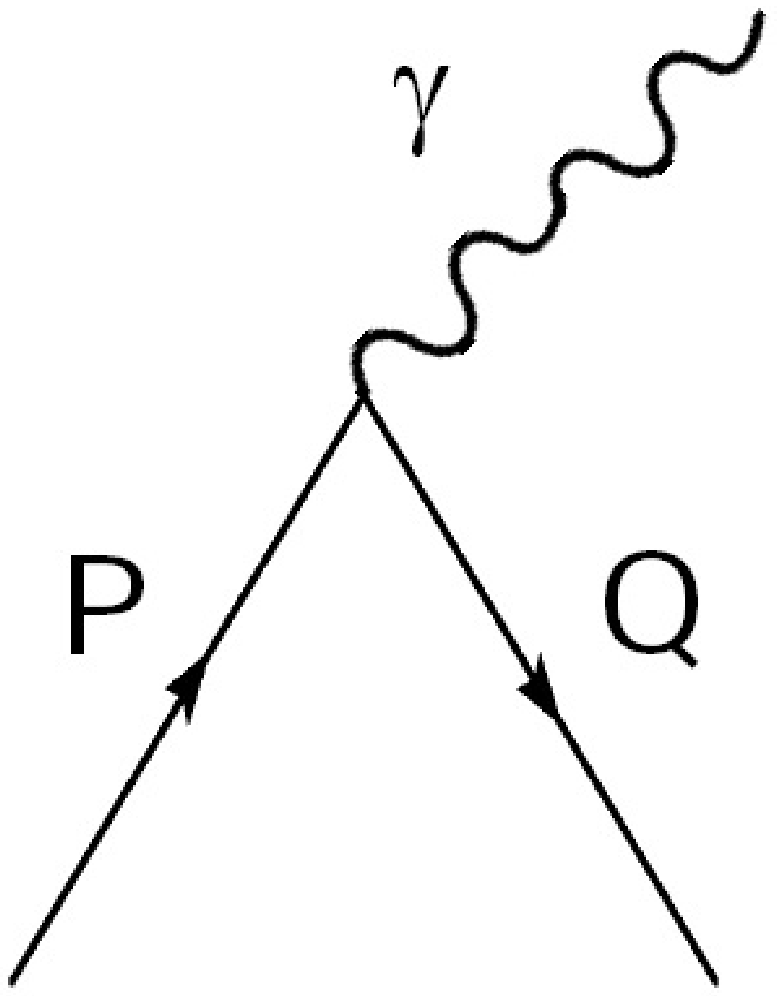}\end{center}

\caption{A partícula progressiva $P$ interage com o fóton $\gamma$ e se transforma
na partícula retrógrada $Q$.}
\label{figura: FD}
\end{figure}

O leitor poderia pensar assim que não há diferenças entre partículas
progressivas e retrógradas, já que aparentemente as últimas são sempre
vistas como às primeiras. No entanto, isso não é bem assim, pois ao
observarmos uma partícula retrógrada como progressiva, algumas de
suas propriedades acabam por se inverterem no processo. Por exemplo,
se a partícula retrógrada tem uma carga elétrica $+e$, então devido
ao princípio de conservação de carga elétrica, devemos observar a
``partícula invertida'' como que portando a carga $-e$. Isso pode
ser mais bem visualizado através da seguinte experiência.

Considere o fenômeno descrito na figura \ref{figura: FD}, onde uma
partícula progressiva $P$, de carga elétrica $+e$, interage com
um fóton $\gamma$ e que, em virtude dessa interação, se torne uma
partícula retrógrada, $Q$. Note que a partícula $Q$ é, em verdade,
a própria partícula $P$ mas que agora se ``movimenta'' para trás
no tempo. Portanto, a carga elétrica da partícula $Q$ é também $+e$.

Entretanto, quando esse processo for fisicamente observado, o observador
fará uso do princípio da inversão (mesmo que de maneira inconsciente)
e o fenômeno passa a ser interpretado da seguinte forma: duas partículas
de igual massa se aproximam e, em dado momento, colidem e se aniquilam,
dando origem a um fóton. Desde que o fóton não tem carga elétrica
e sabemos que a carga elétrica observada da partícula progressiva
é $+e$, decorre que a carga elétrica observada da partícula retrógrada
tem de ser $-e$. A conclusão que se tira disso é que o sinal da carga
elétrica de uma partícula retrógrada se inverte no processo de observação.

Assim, uma partícula retrógrada de massa $m$ e carga $e$ é sempre
vista como uma partícula progressiva de mesma massa e carga elétrica
oposta. Ora, essas são justamente as propriedades esperadas de uma
\emph{antipartícula}. Portanto o princípio da inversão nos permite
concluir que qualquer partícula retrógrada é observada fisicamente
como uma antipartícula. O conceito de antipartículas pode ser visto,
assim, como um conceito puramente relativístico; não é necessário
se falar de mecânica quântica para se introduzir esse conceito%
\footnote{A conexão entre partículas que voltam no tempo com antipartículas
já foi, é claro, proposta por diversos cientistas (\emph{e.g.}, Dirac
\cite{Dirac}, Stückelberg \cite{Stuckelberg,Stuckelberg-1}, Feynman
\cite{Feynman,Feynman-1}, Sudarshan \cite{Sudarshan}, Recami \cite{Recami}
etc.). %
}. 

Por fim, deixe-nos comentar que esses argumentos também são válidos
no caso de partículas do tipo espaço, ou seja, no caso dos táquions.
No entanto, veremos na seção \ref{sec: Dynamics} que no caso dos
táquions a energia se relaciona com o momentum e a massa através da
relação $E^{2}=p^{2}c^{2}-m^{2}c^{4}$, equação esta que descreve
um hiperboloide de folha única. Assim podemos verificar que os táquions
possuem uma propriedade bem interessante: eles podem passar de uma
partícula progressiva a uma retrógrada (e vice-versa) através de um
simples movimento contínuo. Em outras palavras, ao acelerarmos um
táquion podemos fazê-lo se transformar em um antitáquion e vice-versa
(no momento dessa inversão, aliás, o táquion tornar-se-ia uma partícula
momentânea, ou seja, uma partícula de velocidade infinita). Isso,
é claro, só é possível para partículas do tipo espaço.

\section{Dedução das transformações de Lorentz estendidas (em 2 dimensões)\label{sec: Lorentz2}}

Deixe-nos agora mostrar como as transformações de Lorentz podem ser
generalizadas, ou estendidas, à referenciais que se movimentam com
uma velocidade maior que a da luz, bem como para referenciais retrógrados.
Nessa seção nos limitaremos, contudo, a formular uma teoria em duas
dimensões. Como já foi comentado, em quatro dimensões nos deparamos
com diversas dificuldades para se fazer essa extensão, as quais serão
explicadas na seção \ref{sec: Lorentz4}. Apresentaremos a seguir
duas deduções para as \emph{Transformações de Lorentz Estendidas}
(TLE), uma dedução algébrica e outra geométrica.

\textbf{Dedução Algébrica:} desde que, em duas dimensões, os postulados
apresentados na seção anterior são suficientes para mostrar que a
luz se propaga com mesma velocidade $c$ em qualquer referencial inercial,
podemos tomar esse resultado como o nosso ponto de partida. 

Considere assim que um determinado evento tenha as coordenadas $(ct,x)$
em relação a um referencial $R$ e que para um outro referencial inercial
$R'$, que se movimenta em relação a $R$ com velocidade $v$, as
coordenadas desse mesmo evento sejam $(ct',x')$. Suponha além disso
que os eixos coordenados desses referenciais sejam igualmente orientados
e que em $t'=t=0$, a origem dos referenciais coincidam%
\footnote{Daqui em diante, sempre que falarmos nos referenciais $R$ e $R'$
assumiremos implicitamente que a velocidade relativa entre eles é
$v$ e que as condições estabelecidas acima são sempre satisfeitas.%
}.

Nessas condições, se um raio de luz for emitido da origem do referencial
$R$ e no instante $t=0,$ então esse raio de luz se propagará, em
relação a $R$, conforme a equação 
\begin{equation}
x^{2}-c^{2}t^{2}=0,\label{eq:luz}
\end{equation}
e, para $R'$, devido ao princípio de invariância da velocidade da
luz, conforme 
\begin{equation}
x'^{2}-c^{2}t'^{2}=0.\label{eq:luz'}
\end{equation}
(\ref{eq:luz}) e (\ref{eq:luz'}) implica, portanto, que 
\begin{equation}
x'^{2}-c^{2}t'^{2}=\lambda(v)\left(x^{2}-c^{2}t^{2}\right),\label{eq:lambda1}
\end{equation}
onde $\lambda(v)$ não depende das coordenadas e do tempo, mas pode
depender de $v$. 

Por outro lado, desde que a velocidade do referencial $R$ quando
medida por $R'$ é claramente $-v$, segue também que 
\begin{equation}
x^{2}-c^{2}t^{2}=\lambda(-v)\left(x'^{2}-c^{2}t'^{2}\right).\label{eq:lambda2}
\end{equation}
Conclui-se, assim, a partir de (\ref{eq:lambda1}) e (\ref{eq:lambda2}),
que $\lambda(v)\lambda(-v)=1$. Entretanto, a hipótese de que o espaço-tempo
é isotrópico e homogêneo impede que $\lambda\left(v\right)$ possa
depender da direção da velocidade%
\footnote{De fato, é apenas nesse caso que as transformações procuradas formam
um grupo, conferir \cite{Poincare}. %
} e somos levados à condição 
\begin{equation}
\lambda\left(v\right)^{2}=1\quad\Rightarrow\quad\lambda\left(v\right)=\pm1.
\end{equation}

Temos assim dois casos à considerar. Deixe-nos primeiro considerar
o caso em que $\lambda\left(v\right)=+1$. Nesse caso a equação (\ref{eq:lambda1})
se torna 
\begin{equation}
x'^{2}-c^{2}t'^{2}=x^{2}-c^{2}t^{2},\label{eq:lambda+}
\end{equation}
e, como sabemos, a solução de (\ref{eq:lambda+}) nos leva diretamente
às transformações de Lorentz usuais, 
\begin{equation}
ct'=\frac{ct-xv/c}{\sqrt{1-v^{2}/c^{2}}},\quad x'=\frac{x-vt}{\sqrt{1-v^{2}/c^{2}}}.\label{eq:Lorentz1}
\end{equation}
Note que essas transformações contém a identidade (para $v=0$) e
são descontínuas apenas em $v=\pm c$. Consequentemente, essas transformações
devem se aplicar a todo o intervalo $-c<v<c$, e somente a esse intervalo. 

Esperamos assim que no outro caso, isto é, quando $\lambda\left(v\right)=-1$,
as transformações encontradas estejam relacionadas à velocidades maiores
que a da luz. Mostraremos agora que de fato é isso o que ocorre. 

Para $\lambda\left(v\right)=-1$, a equação (\ref{eq:lambda1}) se
torna 
\begin{equation}
x'^{2}-c^{2}t'^{2}=-\left(x^{2}-c^{2}t^{2}\right),\label{eq:lambda-}
\end{equation}
 Através das substituições formais $x\rightarrow\pm i\xi$ e $ct\rightarrow\pm ic\tau$,
podemos reescrever (\ref{eq:lambda-}) como 
\begin{equation}
x'^{2}-c^{2}t'^{2}=\xi^{2}-c^{2}\tau^{2}.\label{eq:xi}
\end{equation}

\begin{figure}[h]
\begin{center}\includegraphics[scale=0.7]{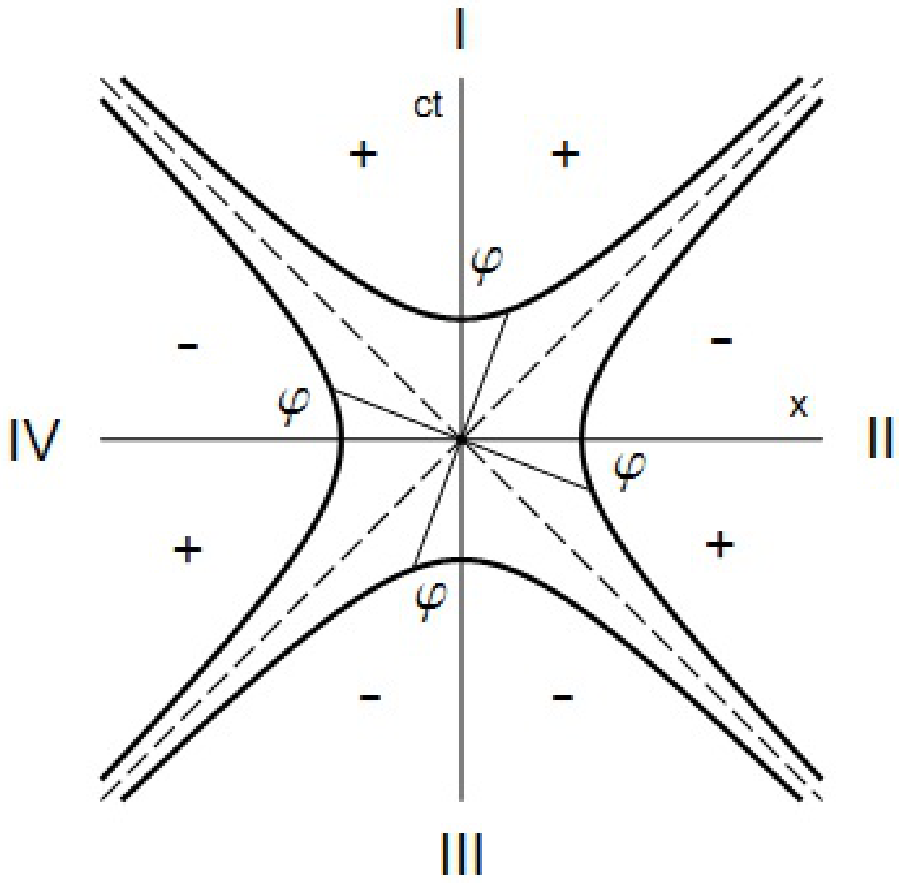}\end{center}\caption{Gráfico da Curva \foreignlanguage{english}{$\left|c^{2}t^{2}-x^{2}\right|=\rho^{2}.$} }
\label{figura: H}
\end{figure}

A equação (\ref{eq:xi}) tem a mesma forma que a equação (\ref{eq:lambda+})
e, portanto, admite mesma solução: 
\begin{equation}
ct'=\frac{c\tau-\xi v/c}{\sqrt{1-v^{2}/c^{2}}},\quad x'=\frac{\xi-v\tau}{\sqrt{1-v^{2}/c^{2}}}.
\end{equation}
 Expressando novamente essas equações em termos de $x$ e $t$, obtemos
\begin{equation}
ct'=\pm\frac{ct-xv/c}{\sqrt{v^{2}/c^{2}-1}},\quad x'=\pm\frac{x-vt}{\sqrt{v^{2}/c^{2}-1}},
\end{equation}
e agora só nos resta remover a ambiguidade dos sinais presente nessas
transformações. O sinal correto, contudo, depende das ``direções de
movimento'' dos referenciais $R$ e $R'$ no espaço-tempo, conforme
pode ser visualizado na figura \ref{figura: H}. No caso de uma transformação
\emph{progressiva} do tipo espaço, facilmente se pode mostrar que
o sinal correto é o \emph{negativo}.

Note que essas equações, assim como no caso anterior, são descontínuas
apenas em $v=\pm c$, mas agora elas são reais somente quando $|v|>c$.
Essas equações representam, por conseguinte, as transformações de
Lorentz entre dois referenciais progressivos cuja velocidade relativa
é maior que a da luz, onde o sinal correto depende se o referencial
$R'$ é progressivo ou retrógrado em relação à $R$ e pode ser visualizado
na figura \ref{figura: H}. 

\textbf{Dedução Geométrica:} do ponto de vista geométrico, as TLE
correspondem a uma rotação (hiperbólica) definida sobre a curva%
\footnote{Tal rotação pode ser mais elegantemente descrita através do conceito
de \emph{números hiperbólicos} \cite{h-numbers}. Um número hiperbólico
é um número da forma $z=a+hb$, com $\left\{ a,b\right\} \in\mathbb{R}$
e $h:\{h^{2}=+1,\: h\notin\mathbb{R}\}$. Definindo-se o conjugado
$\bar{z}=a-hb$, segue que $|\bar{zz}|=|a^{2}-b^{2}|=\rho^{2}$, uma
equação da mesma forma que (\ref{eq:Hip1}). Obtém-se assim uma completa
analogia com os números complexos, mas com a diferença de que agora
esses números descrevem uma geometria hiperbólica. Destacamos também
que o mesmo pode ser feito através da chamada \emph{álgebra geométrica
do espaço-tempo} \cite{Vaz}, com a vantagem de que esse formalismo
talvez permita uma generalização à dimensões maiores.%
} 
\begin{equation}
\left|c^{2}t^{2}-x^{2}\right|=\rho^{2}.\label{eq:Hip1}
\end{equation}
Podemos chamar tal transformação de \emph{rotação hiperbólica estendida.}
Note que (\ref{eq:Hip1}) representa um par de hipérboles equiláteras
 e ortogonais. As duas assíntotas dessa curva dividem o plano em quatro
regiões desconexas, a que chamaremos de regiões I, II, III e IV, respectivamente
(ver figura \ref{figura: H}). 

Para expressar tal rotação, convém introduzir as \emph{funções hiperbólicas
estendidas}, $\cosh\theta$ e $\sinh\theta$, definido-as através
das relações 
\begin{equation}
ct=\rho\cosh\theta,\quad x=\rho\sinh\theta,
\end{equation}
onde $0\leq\theta<2\pi$ é o \emph{parâmetro circular} usual e $\rho$
é dado por (\ref{eq:Hip1}). Note que nessa descrição geométrica a
velocidade passa a ser dada pela expressão 
\begin{equation}
v/c=\tanh\theta.\label{eq:vel}
\end{equation}

As expressões de $\cosh\theta$ e $\sinh\theta$, por sua vez, podem
ser determinadas através das funções hiperbólicas usuais, $\cosh\varphi$
e $\sinh\varphi$, onde $\varphi$ é o parâmetro hiperbólico, $-\infty<\varphi<\infty$.
De fato, se introduzirmos em cada região do plano um parâmetro hiperbólico
$\varphi$ correspondente, o qual deve ser medido, por conveniência,
como mostrado na figura \ref{figura: H}, então as funções $\cosh\varphi$
e $\sinh\varphi$ nos permitem parametrizar cada um dos ramos da curva
(\ref{eq:Hip1}). Assim, uma vez especificada a região na qual $\theta$
pertence, $\rho$ e $\varphi$ determinam de forma única qualquer
ponto da curva (\ref{eq:Hip1}) e, portanto, determinam também as
funções hiperbólicas estendidas. Com essas convenções, encontramos
que 
\begin{equation}
\begin{array}{c}
\cosh\theta\equiv\begin{cases}
+\cosh\varphi, & \theta\in\text{I}\\
-\sinh\varphi, & \theta\in\text{II}\\
-\cosh\varphi, & \theta\in\text{III}\\
+\sinh\varphi, & \theta\in\text{IV}
\end{cases},\\
\\
\sinh\theta\equiv\begin{cases}
+\sinh\varphi, & \theta\in\text{I}\\
+\cosh\varphi, & \theta\in\text{II}\\
-\sinh\varphi, & \theta\in\text{III}\\
-\cosh\varphi, & \theta\in\text{IV}
\end{cases},
\end{array}\label{eq:FH1}
\end{equation}
onde os parâmetros $\theta$ e $\varphi$ relacionam-se através da
fórmula 
\begin{equation}
\tan\theta=\tanh\theta\equiv\begin{cases}
+\tanh\varphi, & \theta\in\left(\text{I, III}\right)\\
-\coth\varphi, & \theta\in\left(\text{II, IV}\right)
\end{cases}.\label{eq:thetaphi}
\end{equation}
As expressões das funções hiperbólicas estendidas também podem ser
encontradas sem se fazer uso das funções hiperbólicas usuais. Para
isso parametrizamos (\ref{eq:Hip1}) através das funções circulares,
colocando $ct=r\cos\theta$ e $x=r\sin\theta$, onde $r=\sqrt{c^{2}t^{2}+x^{2}}$
. Isso nos permite escrever diretamente, 
\begin{equation}
\cosh\theta=\frac{\cos\theta}{\sqrt{\left|\cos2\theta\right|}},\quad\sinh\theta=\frac{\sin\theta}{\sqrt{\left|\cos2\theta\right|}}.\label{eq:FH2}
\end{equation}
ou, em termos da tangente, 
\begin{equation}
\cosh\theta=\frac{\sigma(\theta)}{\sqrt{\left|1-\tan^{2}\theta\right|}},\quad\sinh\theta=\frac{\sigma(\theta)\tan\theta}{\sqrt{\left|1-\tan^{2}\theta\right|}},\label{eq:FH3}
\end{equation}
com 
\begin{equation}
\sigma(\theta)=\begin{cases}
+1, & -\pi/2<\theta<\pi/2\\
-1, & \pi/2<\theta<3\pi/2
\end{cases}.
\end{equation}

A equivalência entre (\ref{eq:FH1}) e (\ref{eq:FH2}) ou (\ref{eq:FH3})
é encontrada quando se leva em conta (\ref{eq:thetaphi}).

Uma vez definidas as funções hiperbólicas estendidas é fácil 
\begin{figure}[h]
\begin{center}\includegraphics[scale=0.5]{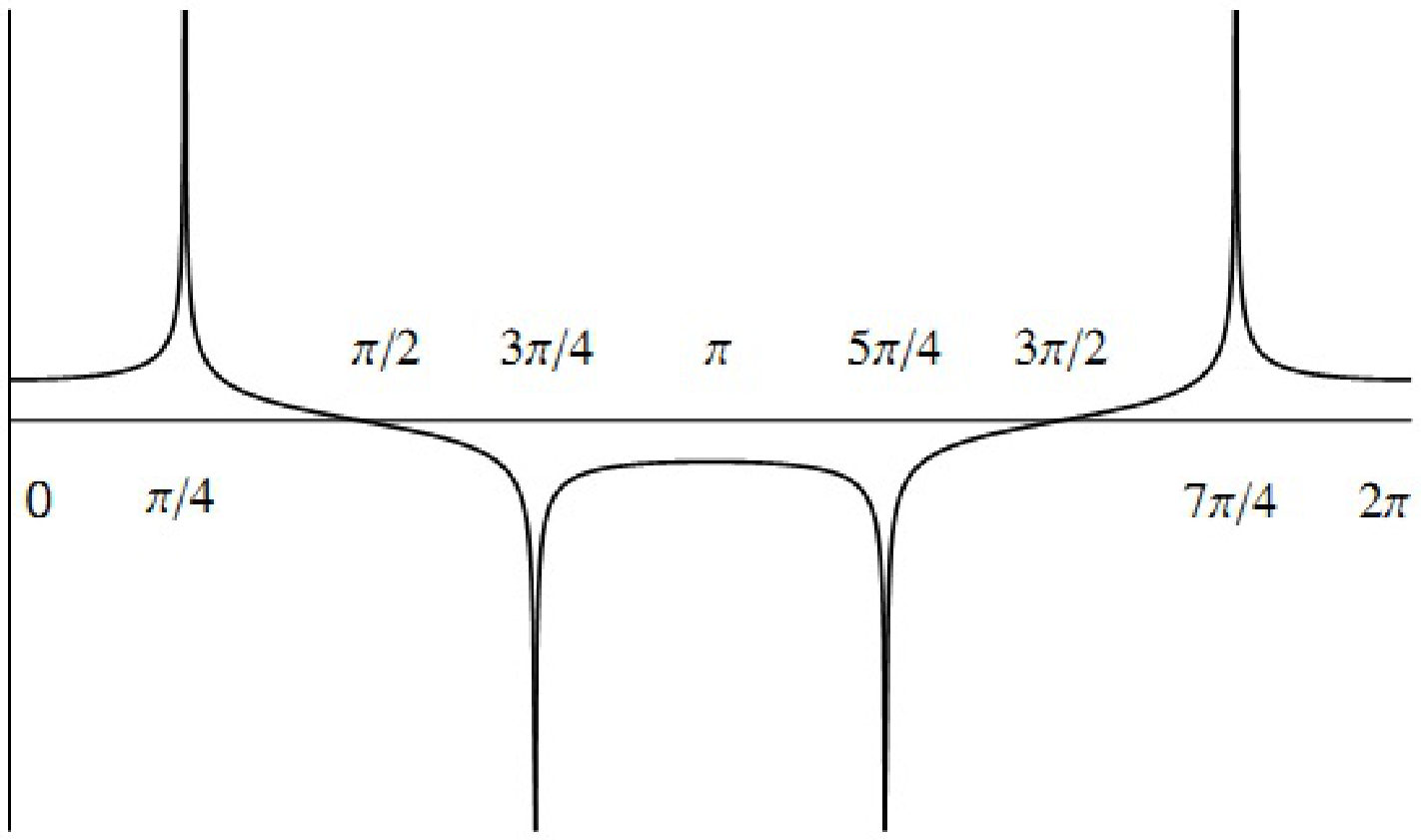}\end{center}\label{figura: C}

\caption{Gráfico da função hiperbólica estendida, $\text{\ensuremath{\cosh\theta}}$.
O gráfico da função $\sinh\theta$ é semelhante a este, mas com uma
defasagem de $\pi/2$ rad.}
\end{figure}
 obter as expressões que descrevem uma rotação hiperbólica estendida.
Sejam $\left(ct,x\right)=\left(\rho\cosh\theta_{1},\rho\sinh\theta_{1}\right)$,
as coordenadas de um ponto do plano para um sistema de coordenadas
$R$, o qual supomos pertencer ao setor I do espaço-tempo. Se efetuarmos
uma rotação hiperbólica passiva (\emph{i.e.}, dos eixos coordenados),
digamos por um ângulo $\theta_{12}$, vamos obter um novo sistema
de coordenadas $R'$ e as coordenadas daquele mesmo ponto passarão
a ser dadas agora por $\left(ct',x'\right)=\left(\rho\cosh\theta_{2},\rho\sinh\theta_{2}\right)$.
Como $\theta_{12}=\theta_{1}-\theta_{2}$, segue que 
\begin{equation}
ct'=\rho\cosh\left(\theta_{1}-\theta_{12}\right),\quad x'=\rho\sinh\left(\theta_{1}-\theta_{12}\right).
\end{equation}

Substituindo as expressões de $\cosh\left(\theta_{1}-\theta_{12}\right)$
e $\sinh\left(\theta_{1}-\theta_{12}\right)$ por qualquer uma das
expressões acima e simplificando as expressões resultantes, levando-se
ainda em conta (\ref{eq:thetaphi}), pode-se verificar que $\left(ct',x'\right)$
relaciona-se com $\left(ct,x\right)$ através das equações 
\begin{equation}
\begin{array}{c}
ct'=\delta\left(\theta_{12}\right)\left(ct\cosh\theta_{12}-x\sinh\theta_{12}\right),\\
\\
x'=\delta\left(\theta_{12}\right)\left(x\cosh\theta_{12}-ct\sinh\theta_{12}\right),
\end{array}
\end{equation}
onde 
\begin{equation}
\delta\left(\theta\right)=\begin{cases}
+1, & \tan^{2}\theta<1\\
-1, & \tan^{2}\theta>1
\end{cases}.\label{eq:delta}
\end{equation}
Por fim, utilizando (\ref{eq:FH3}) e colocando $\tan\theta_{12}=v/c$,
obtemos diretamente as transformações procuradas, que são idênticas
às obtidas algebricamente: 
\begin{equation}
ct'=\varepsilon\left(\theta_{12}\right)\cfrac{ct-xv/c}{\sqrt{\left|1-v^{2}/c^{2}\right|}},\quad x'=\varepsilon\left(\theta_{12}\right)\cfrac{x-vt}{\sqrt{\left|1-v^{2}/c^{2}\right|}},\label{eq:TLE2}
\end{equation}
onde fizemos $\varepsilon\left(\theta_{12}\right)=\sigma\left(\theta_{12}\right)\delta\left(\theta_{12}\right)$.
$\varepsilon\left(\theta_{12}\right)$ determina o sinal que deve
figurar na frente dessas transformações, conforme pode ser visto na
figura \ref{figura: H}.

\section{A lei de composição de velocidades e as transformações inversas\label{sec: velocity}}

As transformações deduzidas na seção anterior formam um grupo, no
qual as transformações de Lorentz ordinárias constituem apenas um
subgrupo. Deixe-nos demonstrar essa estrutura. 

Em primeiro lugar, notemos que a identidade é obtida quando se faz
$v=0$. Mostraremos agora que a composição de duas TLE resulta ainda
em outra TLE. Para isso introduzimos um terceiro referencial inercial
$R''$, que se movimenta em relação a $R'$ com velocidade $u=c\tanh\theta_{23}$.
Por outro lado, supomos que o referencial $R'$ movimenta-se com a
velocidade $v=c\tanh\theta_{12}$ em relação a $R$. A lei de transformação
entre $R$ e $R'$ já conhecemos, e pode ser escrita como 
\begin{equation}
\begin{array}{c}
ct'=\varepsilon\left(\theta_{12}\right)\cfrac{ct-x\tanh\theta_{12}}{\sqrt{\left|1-\tanh^{2}\theta_{12}\right|}},\\
\\
x'=\varepsilon\left(\theta_{12}\right)\cfrac{x-ct\tanh\theta_{12}}{\sqrt{\left|1-\tanh^{2}\theta_{12}\right|}}.
\end{array}\label{eq:C2}
\end{equation}
Por sua vez, a lei de transformação de $R'$ para $R''$ pode ser
escrita por uma expressão análoga: 
\begin{equation}
\begin{array}{c}
ct''=\varepsilon'\left(\theta_{23}\right)\cfrac{ct'-x'\tanh\theta_{23}}{\sqrt{\left|1-\tanh^{2}\theta_{23}\right|}},\\
\\
x''=\varepsilon'\left(\theta_{23}\right)\cfrac{x'-ct'\tanh\theta_{23}}{\sqrt{\left|1-\tanh^{2}\theta_{23}\right|}}.
\end{array}\label{eq:C1}
\end{equation}
Nessas equações, $\varepsilon'\left(\theta_{23}\right)$ determina
os sinais correspondentes às transformações de $R'$ para $R''$,
os quais não precisam necessariamente ser iguais aos sinais presentes
nas transformações de $R$ para $R'$. De fato, enquanto na definição
de $\varepsilon\left(\theta_{12}\right)$ o referencial de partida,
$R$, era suposto pertencer à região I do espaço-tempo, o referencial
$R'$ pode pertencer a qualquer região. Assim, $\varepsilon'\left(\theta_{23}\right)$
é uma função ainda a se determinar.

A substituição de (\ref{eq:C2}) em (\ref{eq:C1}) nos fornece a lei
de transformação entre $R$ e $R''$. Após algumas simplificações,
pode-se verificar que as expressões obtidas possuem a mesma forma
de uma TLE, a saber, 
\begin{equation}
\begin{array}{c}
ct''=\varepsilon''\left(\theta_{13}\right)\cfrac{ct-x\tanh\left(\theta_{13}\right)}{\sqrt{\left|1-\tanh^{2}\left(\theta_{13}\right)\right|}}\\
\\
x''=\varepsilon''\left(\theta_{13}\right)\cfrac{x-ct\tanh\left(\theta_{13}\right)}{\sqrt{\left|1-\tanh^{2}\left(\theta_{13}\right)\right|}}.
\end{array}\label{eq:C3}
\end{equation}
onde, 
\begin{equation}
\varepsilon''\left(\theta_{13}\right)=\frac{\varepsilon\left(\theta_{12}\right)\varepsilon'\left(\theta_{23}\right)}{\delta\left(\theta_{12},\theta_{23}\right)},
\end{equation}
com
\begin{equation}
\delta\left(\theta_{12},\theta_{23}\right)=\begin{cases}
+1, & \tanh\theta_{12}\tanh\theta_{23}<1\\
-1, & \tanh\theta_{12}\tanh\theta_{23}>1
\end{cases},
\end{equation}
e \foreignlanguage{english}{
\begin{equation}
\tanh\left(\theta_{13}\right)=\frac{\tanh\theta_{12}+\tanh\theta_{23}}{1+\tanh\theta_{12}\tanh\theta_{23}}=\tanh\left(\theta_{12}+\theta_{23}\right).\label{eq:comp1}
\end{equation}
}Através de (\ref{eq:thetaphi}) podemos verificar que (\ref{eq:comp1})
consiste em uma generalização da fórmula de adição das tangentes hiperbólicas,
o que revela o seu significado geométrico. Em termos das velocidades,
a equação (\ref{eq:comp1}) pode ser reescrita como 
\begin{equation}
w=\frac{u+v}{1+uv/c^{2}}.\label{eq:comp2}
\end{equation}

A equação (\ref{eq:comp2}) expressa à lei de composição de velocidades
em nossa teoria. Ela é exatamente igual à expressão obtida na teoria
da relatividade usual, mas agora aplica-se a quaisquer valores de
velocidade.

Podemos igualmente mostrar que existe a transformação inversa. Para
isso podemos colocar em (\ref{eq:C3}) as condições $ct''=ct$ e $x''=x$,
e requerer que a transformação resultante seja a identidade. Assim
verificamos que é necessário se ter $\theta_{23}=-\theta_{12}$ e
$\varepsilon''\left(\theta_{13}\right)=1$. Com essas condições torna-se
possível calcular $\varepsilon'\left(\theta_{23}\right)$ através
da expressão resultante de \foreignlanguage{english}{$\varepsilon''\left(\theta_{13}\right)$}.
De fato, encontramos que 
\begin{equation}
\varepsilon'\left(\theta_{23}\right)=\delta\left(\theta_{12}\right)\big/\varepsilon\left(\theta_{12}\right)=\sigma\left(\theta_{12}\right),
\end{equation}
 já que $\delta\left(\theta_{12},-\theta_{12}\right)=\delta\left(\theta_{12}\right)$,
com $\delta\left(\theta_{12}\right)$ definido por (\ref{eq:delta}).

Substituindo esse resultado em (\ref{eq:C1}) obtemos então, em termos
das velocidades, as expressões procuradas das transformações inversas,
\begin{equation}
\begin{array}{c}
ct=\varepsilon^{-1}\left(\theta_{12}\right)\cfrac{ct'+x'v/c}{\sqrt{\left|1-v^{2}/c^{2}\right|}},\\
\\
x=\varepsilon^{-1}\left(\theta_{12}\right)\cfrac{x'+vt'}{\sqrt{\left|1-v^{2}/c^{2}\right|}},
\end{array}\label{eq: inversas}
\end{equation}
onde colocamos $\varepsilon^{-1}\left(\theta_{12}\right)=\sigma\left(\theta_{12}\right)$.
Note que os sinais das transformações inversas diferem dos que aparecem
nas transformações diretas. Essa diferença deve-se ao fato já comentado
de que de que na transformação $R\rightarrow R'$, assumimos que o
referencial de partida pertencia sempre à região I do espaço-tempo,
enquanto que na transformação inversa, $R'\rightarrow R$, é o referencial
de chegada que fixamos à região I. Quando $\left|v\right|<c$ essa
assimetria não interfere em nada, já que nesse caso os sinais são
sempre iguais nas duas definições, contudo, no caso em que $\left|v\right|>c$,
decorre que as transformações inversas não podem ser obtidas simplesmente
pela substituição de $v$ por $-v$, ainda é necessário multiplicá-las
por $-1.$

Por fim, mencionamos que a associatividade das TLE pode ser igualmente
demonstrada, o que completa a sua estrutura de grupo.

\section{Referenciais conjugados\label{sec: conjugated}}

Introduziremos agora um importante conceito que só pode ser contemplado
em uma teoria estendida da relatividade: o conceito de \emph{referenciais
conjugados}. A definição é a seguinte: dois referenciais são ditos
serem conjugados se a velocidade relativa entre eles for infinita.
Assim, se considerarmos o referencial $R',$ que -- lembremos -- se
movimenta com velocidade $v$ em relação a $R$, o referencial conjugado
a $R'$ consiste em um referencial $R^{*}$ cuja velocidade é, em
relação a $R$, igual a $w=c^{2}/v$. De fato, obtemos de (\ref{eq:comp2}),
\begin{equation}
\lim_{u\rightarrow\infty}\left(\frac{u+v}{1+uv/c^{2}}\right)=\frac{c^{2}}{v}.
\end{equation}

A importância do conceito de referenciais conjugados provém do fato
de que uma transformação do tipo espaço entre dois referenciais, digamos,
entre $R$ e $R'$, pode ser obtida por uma transformação de Lorentz
usual entre os referenciais $R$ e $R^{*}$, bastando para isso fazer
as substituições $v\rightleftharpoons c^{2}/v$, \foreignlanguage{english}{$ct^{*}\rightleftharpoons x$}
e \foreignlanguage{english}{$x^{*}\rightleftharpoons ct$}. De fato,
desde que $w=c^{2}/v$ é menor que $c$ para $v>c$, segue que a transformação
de $R$ para $R^{*}$ é dada por \foreignlanguage{english}{
\begin{equation}
ct^{*}=\frac{ct-xw/c}{\sqrt{1-w^{2}/c^{2}}},\quad x^{*}=\frac{x-wt}{\sqrt{1-w^{2}/c^{2}}}.
\end{equation}
}E pode-se verificar que, efetuando-se as substituições indicadas
acima, obtém-se as transformações corretas entre $R$ e $R'$. 

Do ponto de vista geométrico a passagem de um referencial para o seu
conjugado consiste em uma reflexão dos seus eixos coordenados em relação
às assíntotas da curva (\ref{eq:Hip1}), já que tal reflexão tem justamente
o efeito de trocar $ct$ por $x$ e vice-versa (e, consequentemente,
o de substituir $v$ por $c^{2}/v$). Vemos assim que uma transformação
de Lorentz estendida pode ser reduzida a uma transformação de Lorentz
usual através de apropriadas reflexões em torno das assíntotas (para
o caso das transformações do tipo espaço) e em torno da origem (para
uma transformação retrógrada do tipo tempo). Isso nos fornece uma
nova maneira de se deduzir as TLE.

É interessante notar que se uma partícula tem velocidade $u=c^{2}/v$
em relação ao referencial $R$, então ela terá uma velocidade infinita
para o referencial $R'$, ou seja, ela será uma partícula momentânea
para esse referencial. Assim, podemos verificar que se a velocidade
da partícula for $u>c^{2}/v$ e, além disso, tivermos $v<c$, então
essa partícula se tornará retrógrada para $R'$, e será observada
como uma antipartícula por esse referencial. Se, do contrário, tivermos
$v>c$ , então o referencial $R'$ observará uma antipartícula sempre
que \foreignlanguage{english}{$u<c^{2}/v$}. 

Desde que a velocidade relativa do referencial $R'$ em relação a
$R$ é $-v$, decorre também que uma partícula de velocidade $u'=-c^{2}/v$
terá velocidade infinita em $R$. Assim, no caso em que se tem $|v|<c$,
o referencial $R$ verá uma antipartícula se $u'<-c^{2}/v$, e no
caso em que $|v|>c$, ele só observará uma antipartícula se $u'>-c^{2}/v$.
Essas relações, é claro, podem ser mais facilmente encontradas pela
análise (\ref{eq:comp1}) ou (\ref{eq:comp2}).

\section{Réguas e relógios\label{sec: rulers}}

Considere dois relógios idênticos, mas com um deles fixo no referencial
$R$, enquanto que o outro é fixo no referencial $R'$. Desejamos
comparar o ritmo de funcionamento desses relógios, quando medido por
um desses referenciais. Por exemplo, suponha que desejamos comparar
o ritmo do relógio fixo em $R'$ com o relógio fixo em $R$, quando
ambos os relógios são medidos por $R$. Para isso, considere que para
o referencial $R'$, o seu relógio leve o tempo $\tau'$ para dar
uma oscilação completa. O tempo $T$ correspondente a essa oscilação
para o referencial $R$ pode ser encontrado através das transformações
inversas (\ref{eq: inversas}), se lembrarmos que $\Delta x'=0$,
já que o relógio analisado está fixo em $R'$. Obtemos assim que 
\begin{equation}
T=\varepsilon^{-1}\left(\theta\right)\frac{\tau'}{\sqrt{\left|1-v^{2}/c^{2}\right|}}.\label{eq:tempo}
\end{equation}

Podemos então verificar que um relógio progressivo (em relação a $R$)
em movimento trabalha em um ritmo mais lento que um relógio em repouso
quando a sua velocidade for menor que a da luz (como, aliás, é bem
sabido). Mas para um relógio mais rápido que a luz, o seu ritmo de
funcionamento continua a ser menor para $\left|v/c\right|<\sqrt{2}$
e passa a ser maior quando $\left|v/c\right|>\sqrt{2}$. É interessante
notar que para \foreignlanguage{english}{$\left|v/c\right|=\sqrt{2}$}
ambos os relógios voltam a funcionar no mesmo ritmo. Além disso, no
caso de um relógio retrógrado, o relógio em movimento marcará o tempo
de trás para frente, o que expressa o simples fato de que relógios
retrógrados devem marcar o tempo no sentido do futuro para o passado.

Deixe-nos agora verificar como o relógio fixo em $R$ se comporta
em relação ao relógio de $R'$, quando os tempos são medidos por esse
último referencial. Nesse caso devemos utilizar as transformações
diretas e obtemos 
\begin{equation}
T'=\varepsilon\left(\theta\right)\frac{\tau}{\sqrt{\left|1-v^{2}/c^{2}\right|}},\label{eq:tempo2}
\end{equation}
onde $\tau$ é o tempo gasto pelo relógio fixo em $R$ (\emph{i.e.},
que se move com velocidade $-v$ para $R'$) para dar uma oscilação
completa, quando esse tempo é medido por $R$. Agora, o sinal que
aparece em (\ref{eq:tempo2}) é determinado conforme a figura (\ref{figura: H})
e a análise se torna pouco mais ou menos complicada. É claro que não
há problemas quando $\left|v\right|<c$, passemos então a analisar
o caso em que $\left|v\right|>c$. Suponha primeiro que o referencial
$R'$ seja progressivo em relação a $R$. Nesse caso encontramos que
$\varepsilon\left(\theta\right)=-1$ e o relógio que para $R'$ se
movimenta com velocidade $-v$ trabalha no sentido contrário ao do
relógio fixo em $R'$. Isso significa que o relógio de $R$ é retrógrado
em relação à $R'$. Que isso é verdade pode ser verificado através
do que foi comentado na seção anterior, onde se deve notar que $u=0$
e $\left|v\right|>c$ (e que, por conseguinte, $u<c^{2}/v$). 

No entanto, essa é uma situação interessante porque acabamos de ver
que para o referencial $R$, o relógio em movimento funciona no mesmo
sentido que o seu relógio, em repouso. Assim, enquanto um dos referenciais
insiste que os relógios funcionam ao contrário, o outro discorda,
afirmando que ambos os relógios trabalham no mesmo sentido! No caso
em que o referencial $R'$ é retrógrado em relação a $R$ a assimetria
ainda persiste, mas agora é o referencial $R$ que verá ambos os relógios
funcionarem no sentido contrário, enquanto que para $R'$ os relógios
funcionarão no mesmo sentido. Essa assimetria, é claro, tem origem
na própria assimetria entre as transformações diretas e inversas.

Considere agora duas réguas idênticas, uma colocada no referencial
$R$ e outra no referencial $R'$. Desejamos comparar o comprimento
dessas réguas, quando analisadas por um desses referenciais. Se $l_{0}'$
é o comprimento da régua fixa em $R'$ quando medida por esse referencial,
então o seu comprimento $L$, como medido por $R$, é obtido ao se
determinar onde se encontram os estremos da régua em movimento em
um dado instante $t$. Fazendo uso das transformações diretas, encontramos
que 
\begin{equation}
L=\varepsilon\left(\theta\right)\cdot l_{0}'\sqrt{\left|1-v^{2}/c^{2}\right|}.
\end{equation}

Para $\left|v\right|<c$ reencontramos à contração de Lorentz, mas
para $\left|v\right|>c$ obtemos que a régua em movimento será menor
que a régua em repouso quando $\left|v/c\right|<\sqrt{2}$ e voltará
a ter o mesmo comprimento quando $\left|v/c\right|=\sqrt{2}$. Por
fim, para $\left|v/c\right|>\sqrt{2}$ teremos uma ``dilatação de
Lorentz.'' Além disso, considerando que o referencial $R'$ é progressivo
em relação a $R$, decorre que para $R$ a régua em movimento estará
orientada contrariamente a sua régua, em repouso.

Se, por outro lado, as medições são feitas por $R'$, então encontramos
que 
\begin{equation}
L'=\varepsilon^{-1}\left(\theta\right)\cdot l_{0}\sqrt{\left|1-v^{2}/c^{2}\right|},
\end{equation}
e agora para $R'$ a régua em movimento (que tem velocidade $-v$)
apontará para o mesmo sentido que a sua régua em repouso. Encontramos
assim as mesmas assimetrias comentadas mais acima. 

Esses resultados podem, é claro, ser obtidos -- e mais bem compreendidos
-- através dos diagramas de Minkowski.

\section{Dinâmica\label{sec: Dynamics}}

Nesta seção pretendemos responder a algumas questões referentes à
dinâmica dos táquions. As expressões para a energia e o momentum de
uma partícula mais rápida que a luz serão deduzidas e mostraremos
como tais partículas se comportam quando na presença de um campo de
forças.

Como ponto de partida assumiremos que o princípio da ação estacionária
aplica-se também à partículas mais rápidas que a luz. Isso, é claro,
segue diretamente da hipótese de que o espaço-tempo é isotrópico e
homogêneo, uma vez que sabemos que esse princípio é válido para partículas
mais lentas que a luz. 

Como se sabe, o princípio da ação estacionária afirma que existe uma
quantidade $S$, chamada \emph{ação}, que assume um valor extremo
(máximo ou mínimo) para qualquer movimento possível de um sistema
mecânico (no nosso caso, uma partícula). Por outro lado, na ausência
de forças o movimento de uma partícula corresponde a uma geodésica
do espaço-tempo, a qual se reduz a uma reta quando desprezamos a gravitação.
Isso significa que no caso de uma partícula livre a diferencial da
ação $dS$ deve ser proporcional ao elemento de linha $ds$ da partícula
e podemos escrever 
\begin{equation}
dS=\alpha ds,\quad ds=\sqrt{\left|c^{2}dt^{2}-dx^{2}-dy^{2}-dz^{2}\right|}.\label{eq:S}
\end{equation}
Temos de enfatizar, entretanto, que a constante de proporcionalidade
$\alpha$ pode assumir valores diferentes em regiões diferentes do
espaço-tempo, uma vez que essas regiões são completamente desconexas.
Portanto, é conveniente considerar cada caso separadamente. 

No caso de uma partícula progressiva do tipo tempo, (\ref{eq:S})
assume a forma, \foreignlanguage{english}{
\begin{equation}
dS=\mathcal{L}\left(u\right)dt,\quad\mathcal{L}\left(u\right)=\alpha c\sqrt{1-u^{2}/c^{2}},\label{eq:L1}
\end{equation}
}onde introduzimos a função de LaGrange, \foreignlanguage{english}{$\mathcal{L}\left(u\right)$},
para expressar a ação em termos da velocidade da partícula. 

Como se sabe, as expressões para a energia e momentum são obtidas
pelas fórmulas 
\begin{equation}
p\left(u\right)=\frac{\partial\mathcal{L}\left(u\right)}{\partial u},\quad E\left(u\right)=u\left[\frac{\partial\mathcal{L}\left(u\right)}{\partial u}\right]-\mathcal{L}\left(u\right).\label{eq:pE}
\end{equation}
Aplicando-se (\ref{eq:pE}) em (\ref{eq:L1}) obtemos, assim, 
\begin{equation}
p\left(u\right)=-\frac{\alpha u/c}{\sqrt{1-u^{2}/c^{2}}},\quad E\left(u\right)=-\frac{\alpha c}{\sqrt{1-u^{2}/c^{2}}}.
\end{equation}
Para determinar $\alpha$ podemos usar o fato de que essas expressões
devem, para baixas velocidades, se reduzir às expressões obtidas pela
mecânica de Newton. Assim, por exemplo, se expandirmos a expressão
para o momentum em uma série de potências de $u/c$ e retermos apenas
o primeiro termo, encontramos que $p\approx-\alpha u/c$, ao passo
que a mecânica de Newton nos fornece $p=mu$. Assim obtemos $\alpha=-mc$
e ficamos com 
\begin{equation}
p\left(u\right)=\frac{mu}{\sqrt{1-u^{2}/c^{2}}},\quad E\left(u\right)=\frac{mc^{2}}{\sqrt{1-u^{2}/c^{2}}},
\end{equation}
 que são justamente as expressões conhecidas da relatividade usual. 

Já caso de uma partícula progressiva do tipo espaço (\emph{i.e}.,
no caso de um táquion progressivo), a função de LaGrange assume a
forma 
\begin{equation}
\mathcal{L}\left(u\right)=\alpha c\sqrt{u^{2}/c^{2}-1}.
\end{equation}
E obtemos, por meio de (\ref{eq:pE}), as seguintes expressões para
o momentum e energia dessa partícula, 
\begin{equation}
p\left(u\right)=\frac{\alpha u/c}{\sqrt{u^{2}/c^{2}-1}},\quad E\left(u\right)=\frac{\alpha c}{\sqrt{u^{2}/c^{2}-1}}.
\end{equation}
A constante $\alpha$, entretanto, não pode mais ser calculada através
da comparação dessas expressões com as que são obtidas na mecânica
de Newton, já que a velocidade da partícula é nesse caso sempre maior
que a da luz. Mas podemos, em contrapartida, calcular o limite dessas
expressões para quando $u\rightarrow\infty$, e assim encontramos
que 
\begin{equation}
\lim_{u\rightarrow\infty}p\left(u\right)=\alpha,\quad\lim_{u\rightarrow\infty}E\left(u\right)=0,
\end{equation}
de onde se pode ver que $\alpha$ é igual ao momentum de uma partícula
momentânea, ou seja, de velocidade infinita. 

\begin{figure}[h]
\begin{center}\includegraphics[scale=0.5]{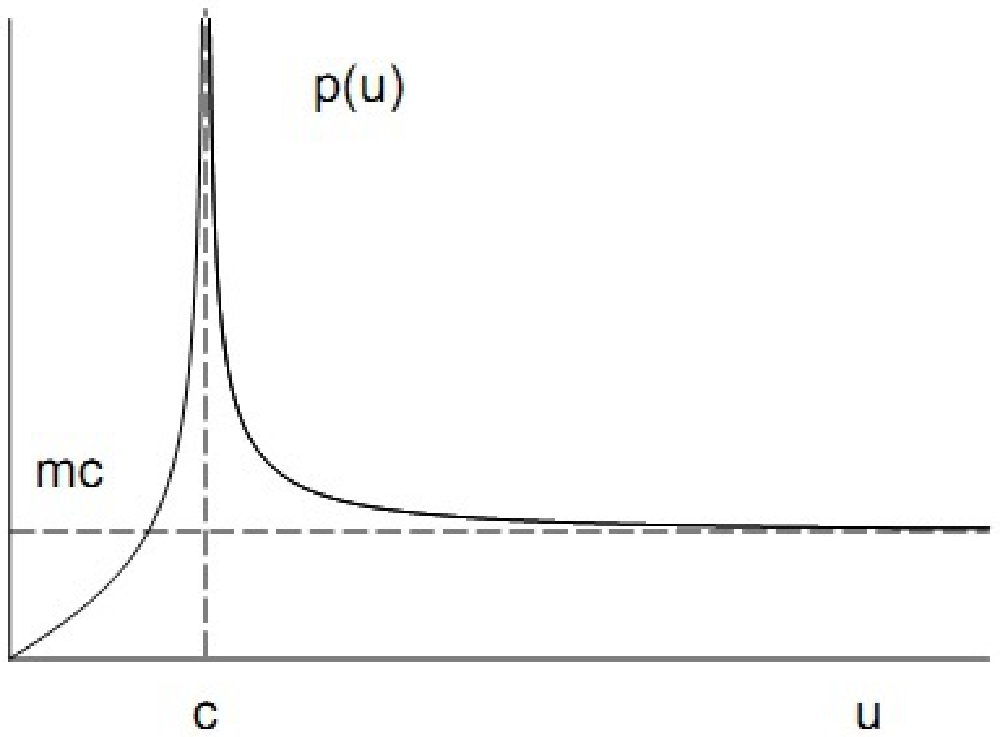} \includegraphics[scale=0.5]{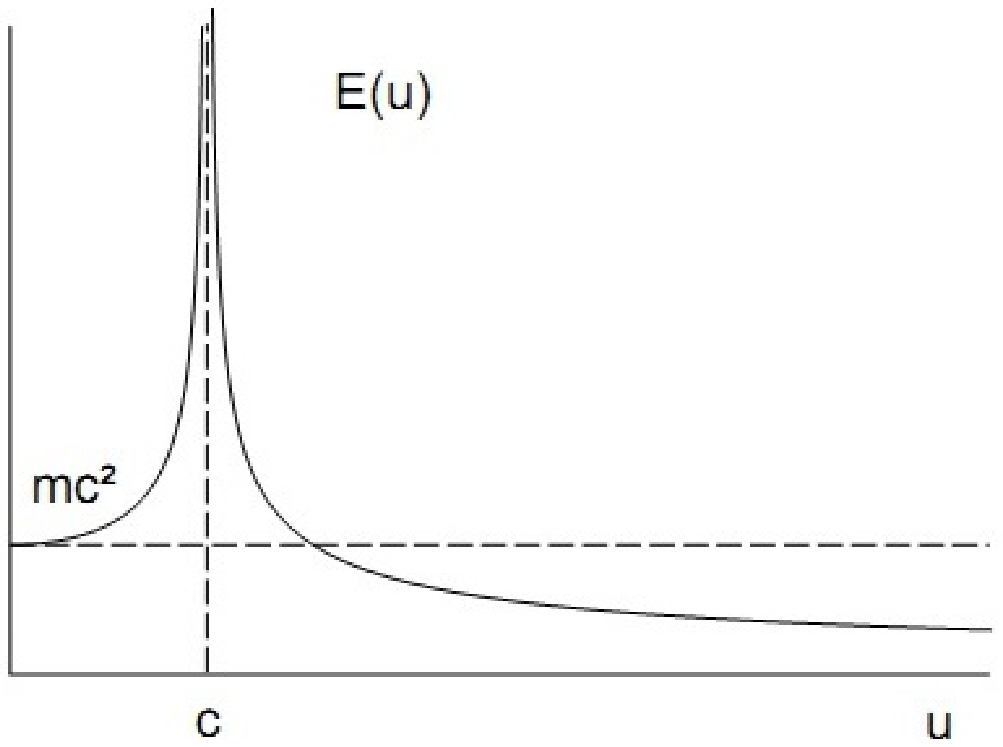}\end{center}\caption{Gráficos para o momentum e energia de uma partícula progressiva com
velocidade $0\leq u<\infty$.}

\end{figure}

Desde que a massa de uma partícula deve ser um invariante universal,
segue que podemos definir uma métrica também no espaço da energia
e dos momenta. Fazendo uma analogia com $(\ref{eq:metrica})$, podemos
definir essa \emph{métrica dinâmica} como 
\begin{equation}
c^{2}dm=\sqrt{\left|dE^{2}-c^{2}dp_{x}^{2}-c^{2}dp_{y}^{2}-c^{2}dp_{z}^{2}\right|}.
\end{equation}
Assim, segue que a massa, a energia e o momentum de uma partícula
devem estar relacionados sempre pela fórmula $mc^{2}=\sqrt{\left|E^{2}-c^{2}p^{2}\right|}$.
Se colocarmos, nessa última equação, $E=0$, obtemos que $p=mc=\alpha$
(onde assumimos $p$ é positivo para $u$ positivo). E assim concluímos
que o momentum e a energia de um táquion são dadas pelas as expressões
\begin{equation}
p\left(u\right)=\frac{mu}{\sqrt{u^{2}/c^{2}-1}},\quad E\left(u\right)=\frac{mc^{2}}{\sqrt{u^{2}/c^{2}-1}}.\label{eq:pE2}
\end{equation}
Esse resultado pode também ser obtido através do emprego dos referenciais
conjugados, introduzidos na seção \ref{sec: conjugated}. Suponha
que uma partícula tenha velocidade $u>c$ para um referencial $R$.
Então para o referencial conjugado $R^{*}$ ela terá energia $E^{*}=cp$
e momentum $p^{*}=E/c$, enquanto que a sua velocidade passa a ser
$w=c^{2}/u$, que é menor que $c$. Mas as expressões para o momentum
e energia de um brádion (\emph{i.e}., de uma partícula do tipo tempo)
são dadas por (\ref{eq:pE}), e obtemos, por conseguinte, 
\begin{equation}
p^{*}=\frac{mw}{\sqrt{1-w^{2}/c^{2}}},\quad E^{*}=\frac{mc^{2}}{\sqrt{1-w^{2}/c^{2}}}.
\end{equation}
Podemos assim verificar que as substituições de $E^{*}$, $p^{*}$
e $w$ nas fórmulas acima resultam exatamente na expressão (\ref{eq:pE2}).

No caso de partículas retrógradas, a função de LaGrange troca de sinal,
já que $dt$ é negativo e $\mathcal{L}=dS/dt$. Por conseguinte, partículas
retrógradas devem ter energia negativa, um resultado que já havíamos
indicado na seção \ref{sec: antiparticles}.

Por fim, deixe-nos mostrar como uma partícula do tipo espaço deve
se comportar quando submetida a um campo de forças. A força que atua
na partícula continua, é claro, sendo dada pela lei de Newton, $\dot{p}=F$,
mas agora $p$ é dado por (\ref{eq:pE2}). Calculando a derivada,
obtemos 
\begin{equation}
F=-\frac{ma}{\left(u^{2}/c^{2}-1\right)^{3/2}},
\end{equation}
onde $a=du/dt$ denota a aceleração da partícula. Note que a força
é dirigida no sentido oposto à aceleração. Consequentemente duas partículas
do tipo espaço que \emph{dinamicamente} se atraem (se repelem) devem
se afastar (se aproximar) uma da outra. Os conceitos de atração-aproximação
e de repulsão-afastamento, portanto, não são mais equivalentes quando
lidamos com táquions. De fato, se tentássemos se aproximar (ou se
afastar) de um táquion, a sua velocidade relativa diminuiria (aumentaria),
um efeito que é contrário ao nosso senso comum mas que pode ser demonstrado
por uma simples análise de (\ref{eq:comp2}).

\section{Causalidade e o paradoxo de Tolman\label{sec: Tolman}}

Um dos argumentos geralmente empregados para se mostrar que partículas
mais rápidas que a luz não podem existir é que a sua existência implicaria
uma violação do princípio de causalidade. Com efeito, suponha que
um táquion seja emitido por um corpo $A$ no instante $t_{A}$ e que
ele seja absorvido por outro corpo $B$ no instante $t_{B}$, onde
$t_{B}>t_{A}$. Como esses dois eventos (emissão de um táquion por
$A$ e sua absorção por $B$) estão separados por uma distância do
tipo espaço, sabemos da teoria da relatividade que é possível encontrar
um outro referencial, com velocidade menor que a da luz, onde a ordem
cronológica desses eventos se inverte. Assim, se assumirmos que o
evento em $A$ é a causa do evento $B$, então chegaremos a conclusão
de que, para esse referencial em movimento, o efeito precede a sua
causa. 

Portanto, se admitirmos que os conceitos de causa e efeito são \emph{absolutos},
e que a causa sempre precede o seu efeito, então efetivamente somos
levados a conclusão de que partículas mais rápidas que a luz não podem
existir. No entanto, não há nada do ponto de vista matemático que
dê suporte a essa hipótese. Pelo contrário, se admitirmos uma completa
isotropia e homogeneidade do espaço e tempo, então não temos outra
escolha senão considerar os conceitos de causa e efeito como conceitos
\emph{relativos}. 

Já vimos, é claro, outras quantidades que eram consideradas absolutas
na teoria da relatividade usual e passaram agora a serem vistas como
que relativas. Um exemplo é o conceito de emissão e absorção. Com
efeito, uma vez que pelo princípio da inversão não observamos partículas
retrógradas mas sim antipartículas progressivas, os processos de emissão
(absorção) de partículas retrógradas são sempre observadas como uma
absorção (emissão) de antipartículas. Podemos, é claro, descobrir
se uma partícula foi de fato emitida (absorvida) se analisarmos o
processo no referencial onde a fonte (emissor) está em repouso, nesse
caso falaremos de uma emissão (absorção) \emph{intrínseca} ou \emph{própria}. 

O caráter relativo da causalidade, todavia, não leva a nenhuma contradição
à teoria. Na natureza, um fenômeno nunca descreve um evento isolado,
mas sim uma sucessão contínua de eventos, ou seja, um \emph{processo}
físico. Geometricamente um processo descreve uma curva contínua no
espaço-tempo e portanto tem um caráter absoluto: a ordem temporal
com que os eventos ocorrem podem diferir de referencial para referencial,
mas a curva em si, que corresponde ao fenômeno em questão, é a mesma
para qualquer um deles. Isso, por exemplo, é suficiente para mostrar
que não podemos voltar no tempo e matar o nosso avô. De fato, o simples
fato de existirmos significa que existe uma curva no espaço tempo
conectando o nosso avô a nós e, por essa curva ter caráter absoluto,
ela não pode ser (estar) desconectada por (para) qualquer observador,
mesmo que este voltasse no tempo.

\begin{figure}[h]
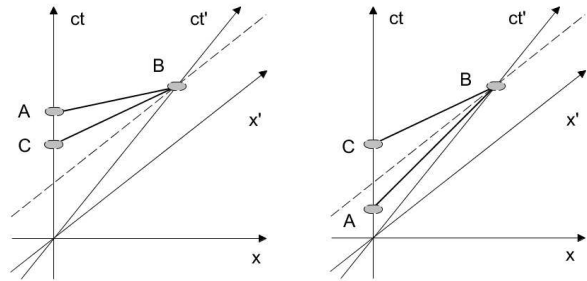

\begin{center}

\includegraphics[scale=0.35]{T1} \hspace{0.2cm}\includegraphics[scale=0.35]{T2}

\end{center}

\caption{Experimento imaginado por Tolman.}

\label{figura: T}
\end{figure}

Como exemplo de um paradoxo que envolve o conceito de causalidade,
deixe-nos analisar o interessante paradoxo proposto e discutido por
Tolman \cite{Tolman} em 1917, embora já tivesse sido apontado dez
anos antes por Einstein em \cite{Einstein}. Maiores detalhes podem
ser conferidos em \cite{Bohm,Recami-1,Recami-2}.

O paradoxo consiste em mostrar que se existissem partículas mais rápidas
que a luz, então seria possível enviar informações para o passado.
Para mostrar isso, considere que os referenciais $R$ e $R'$ (com
$R'$ movendo-se com uma velocidade $v<c$ em relação a $R$) sejam
equipados com certos aparelhos telefônicos especiais, cuja comunicação
é feita através da transmissão de táquions. Imagine, assim, que um
observador em $R$ envie (evento A) uma mensagem, digamos na forma
de uma pergunta, para $R'$ e que, após o seu recebimento (evento
B), um observador em $R'$ mande de volta para $R$ a sua resposta,
que é então recebida por $R$ (evento C). Acontece que, como pode
ser visto na figura \ref{figura: T}a, o observador em $R'$ pode
calibrar o seu aparelho de modo que a mensagem enviada por ele chegue
em $R$ antes que esse observador tivesse enviado a sua pergunta.
Assim, uma vez que $R$ obtém a resposta antes mesmo de fazer a pergunta,
$R$ poderia deixar de fazê-la, mas caso ele tome essa decisão, também
não poderia haver qualquer resposta por parte de $R'$, o que constitui
uma situação paradoxal. 

A falha nessa argumentação é que ela mistura as descrições dos referenciais
quando os termos ``emissão'' e ``absorção'' são empregados. De fato,
mostraremos agora (seguindo \cite{Recami}) que se o referencial $R'$
receber \emph{intrinsecamente} a pergunta enviada por $R$, e também
enviar \emph{intrinsecamente} a sua resposta, então o referencial
$R$ sempre receberá a resposta de $R'$ depois de ter enviado a sua
pergunta.

Para isso, suponha que a pergunta de $R$ é enviada com uma velocidade
$u_{1}$, e que a resposta volte com uma velocidade $u_{2}$, com
ambas as velocidades maiores que a da luz e medidas por $R$. Então
note que para a resposta de $R'$ chegar antes da pergunta de $R$,
é necessário que se tenha $\left|u_{1}\right|>\left|u_{2}\right|$,
porque para $R$ o evento $C$ é anterior a $A$ e ambos chegam em
$B$ no mesmo instante. No entanto, para que o referencial $R'$ receba
\emph{intrinsecamente} a pergunta de $R$ é necessário que se tenha
$\left|u_{1}\right|<c^{2}/v$, já que do contrário, conforme vimos
na seção \ref{sec: conjugated}, essa mensagem seria retrógrada para
$R'$ e o observador em $R'$ veria simplesmente a emissão de uma
mensagem por meio da transmissão de antitáquions, conforme prediz
o princípio da inversão. Da mesma forma, para que $R'$ emita\emph{
intrinsecamente }a sua resposta, devemos ter $\left|u_{2}\right|>c^{2}/v$,
já que a resposta é enviada na direção negativa do eixo $X$ de $R$.
Assim, mostramos que \foreignlanguage{english}{$\left|u_{2}\right|>\left|u_{1}\right|$},
uma contradição. 

Portanto, se o referencial $R'$ de fato recebe a pergunta de $R$
e de fato emite a sua resposta, então o processo será descrito, em
verdade, pela figura \ref{figura: T}b e não pela figura \ref{figura: T}a.
Nesse caso, a resposta sempre chega depois de a pergunta ser enviada
pelo referencial $R$.

É claro que a figura \ref{figura: T}a também representa um processo
físico válido, mas não se refere efetivamente à situação descrita
no paradoxo de Tolman. Além disso, cada referencial terá a sua própria
visão do que ocorre. Por exemplo, para o referencial $R$ a figura
\ref{figura: T}a corresponde ao caso onde, primeiramente, há o envio
da mensagem $C$ (por meio de antitáquions) e, logo depois, o envio
da mensagem $A$ (por meio de táquions), por fim, ambas as mensagens
chegam juntas em $B$. Já para $R'$ tudo se passa como se ele enviasse
a mensagem $C$ para $R$ por meio de antitáquions e, ao mesmo tempo,
enviasse a mensagem $A$ por meio de táquions, então após algum tempo
ele veria que o observador em $R$ recebe a mensagem $C$ e depois
a mensagem $A$. Note que a ordem cronológica dos eventos depende
do referencial considerado, mas o processo em si (que interliga os
eventos $A$, $B$ e $C$) é único.

\section{Dificuldades para se formular a teoria dos táquions em 4 dimensões
\label{sec: Lorentz4}}

Na seção \ref{sec: Lorentz2} conseguimos com êxito mostrar que em
duas dimensões as transformações de Lorentz podem ser estendidas,
de modo a relacionar qualquer par possível e imaginável de referenciais
inerciais. Quando, porém, tentamos fazer o mesmo em quatro dimensões,
nos deparamos com sérias dificuldades, as quais serão brevemente discutidas
aqui.

Deixe-nos lembrar o leitor que ao estender as transformações de Lorentz
em duas dimensões, fizemos o uso de um importante princípio: o que
afirma que a velocidade da luz é a mesma em qualquer referencial inercial.
Vimos, aliás, que esse princípio pode ser deduzido a partir dos postulados
apresentados na seção \ref{sec: space-time}, de modo que ele pode
ser visto como um teorema da própria teoria da relatividade. Quando,
todavia, passamos para um universo de 4 dimensões, não é mais possível
mostrar a validade geral desse teorema da mesma forma que fizemos
lá, de modo que em quatro dimensões não podemos nem mesmo afirmar
que a luz se propaga em superfícies esféricas e com velocidade $c$
em \emph{qualquer }referencial inercial.

Para mostrar mais claramente a origem dessa dificuldade, deixe-nos
considerar por exemplo uma fonte de luz fixa na origem do referencial
$R$ e supor que em relação a esse referencial a luz se propague em
superfícies esféricas e com velocidade $c$. Em quatro dimensões as
assíntotas do espaço-tempo são substituídas por um cone -- o cone
de luz. A região do espaço-tempo ocupada por um raio de luz que é
emitida pela fonte fixa em $R$ sempre pertence a esse cone de luz. 

A teoria da relatividade usual mostra também que para qualquer outro
referencial inercial do tipo tempo, a propagação dessa luz é ainda
representada por esse mesmo cone de luz. Mas como a luz deve se comportar
para um referencial $R'$ do tipo espaço? Nesse caso, o eixo do tempo
desse referencial estará sempre na região externa ao cone de luz de
$R$, de modo que não é evidente que a luz emitida pela fonte fixa
em $R$ deva se propagar em superfícies esféricas de velocidade $c$
para $R'$. Além disso, podemos nos perguntar o que acontece quando
uma fonte fixa em $R'$ emite luz. Será que para $R'$ essa luz se
propagará em superfícies esféricas de velocidade $c$? E para o referencial
$R$, onde a fonte tem velocidade $v>c$, o que ocorre?

Como se pode ver, há várias questões que não podem ser respondidas
de imediato, sem se ter informações adicionais. De fato, o comportamento
da luz depende de suas propriedades intrínsecas, as quais são determinadas
por uma teoria eletromagnética que \emph{a priori} desconhecemos.
Deixe-nos analisar um pouco mais essas possibilidades.

Podemos supor, por exemplo, que a fonte de luz fixa em $R'$ (que
vamos supor ter velocidade maior que a da luz) também emita ondas
esféricas e com velocidade $c$ para esse mesmo referencial. Então
agora deve existir um outro cone de luz, correspondente à esse referencial
do tipo espaço e que deve contornar o eixo do tempo de $R'$. Uma
vez que o eixo do tempo de $R$ sempre estará fora desse novo cone
de luz, segue que para $R$ a luz emitida pela fonte em movimento
não se propagará em superfícies esféricas de velocidade $c$, mas
antes terá a forma de um hiperboloide de dupla face. Nesse caso a
velocidade dos raios luminosos dependeria, por conseguinte, da sua
direção de propagação e o espaço-tempo não poderia mais ser considerado
isotrópico.

Se, do contrário, considerarmos que a luz emitida pela fonte em $R'$
se propaga em superfícies esféricas com respeito ao referencial $R$,
ou seja, que no universo existe apenas um cone de luz, e que a luz
sempre ``percorre'' esse cone, não importando qual seja a velocidade
da fonte que a emitiu, então podemos verificar que embora a luz se
propague em superfícies esféricas de velocidade $c$ para $R$, o
mesmo não ocorre para $R'$, que observa a luz se propagar em hiperboloides
de dupla face.

Para que a luz se propague em superfícies esféricas de velocidade
$c$ em ambos os referenciais é necessário que as coordenadas transversais,
$y'$ e $z'$, do referencial $R'$ sejam imaginárias (assumindo que
$y$ e $z$ são reais). Pois somente desse modo se pode mapear um
cone de luz do ``tipo tempo'' em um cone de luz do ``tipo espaço.''
Com efeito, isso pode ser mostrado ao se deduzir as TLE em 4 dimensões,
assumindo desde o início a validade do princípio de invariância da
velocidade da luz para qualquer referencial inercial. De fato, teremos
nesse caso, 
\begin{equation}
x'^{2}+y'^{2}+z'^{2}-c^{2}t'^{2}=\lambda\left(v\right)\left(x^{2}+y^{2}+z^{2}-c^{2}t^{2}\right),
\end{equation}
no lugar de (\ref{eq:lambda1}). Como se sabe, a solução dessa equação
para $\lambda\left(v\right)=+1$ consiste nas transformações de Lorentz
usuais, enquanto que, para $\lambda\left(v\right)=-1$ obtemos 
\begin{equation}
\begin{array}{c}
ct'=\pm\cfrac{ct-xv/c}{\sqrt{v^{2}/c^{2}-1}},\quad x'=\pm\cfrac{x-vt}{\sqrt{v^{2}/c^{2}-1}},\\
\\
y'=\pm iy,\quad z'=\pm iz,
\end{array}\label{eq:TLE4D}
\end{equation}
e as coordenadas transversais $y'$ e $z'$ tornam-se imaginárias.
A introdução de coordenadas imaginárias é, no entanto, desprovida
de qualquer sentido físico%
\footnote{Alguns autores, como por exemplo em \cite{Dawe}, argumentam que a
introdução de coordenadas imaginárias na teoria não constitui um problema.
A solução apresentada por eles consiste em considerar os próprios
eixos coordenados $Y'$ e $Z'$ como imaginários, de modo que os observadores
do referencial (do tipo espaço) $R'$ sempre interpretem as coordenadas
$y'$ e $x'$ como \emph{relativamente} reais. Isso, todavia, não
é consistente. De fato, considere por exemplo um raio de luz que se
propague na direção $Y$ do referencial $R$. Para o referencial $R'$,
a luz caminha por uma trajetória oblíqua, e as componentes de sua
velocidade terão as magnitudes $\left|c_{x}'\right|=v$ e $\left|c_{y}'\right|=c\sqrt{v^{2}/c^{2}-1}$.
Assim, para que o raio de luz se propague com a velocidade $c'=c$,
é estritamente necessário que o quadrado de $c_{y}'$ seja um número
negativo, o que contraria a hipótese de que $c_{y}'$ é um número
\emph{relativamente} real.%
}. 

Se, por sua vez, negarmos a validade do princípio de invariância da
velocidade da luz, então obtemos as transformações
\begin{equation}
\begin{array}{c}
ct'=\pm\cfrac{ct-xv/c}{\sqrt{v^{2}/c^{2}-1}},\quad x'=\pm\cfrac{x-vt}{\sqrt{v^{2}/c^{2}-1}},\\
\\
y'=\pm y,\quad z'=\pm z.
\end{array}
\end{equation}
em que todas as coordenadas são reais. O problema é que agora estamos
nos casos discutidos anteriormente, onde o espaço-tempo não pode mais
ser considerado isotrópico. 

Note que em cada uma das possibilidades discutidas acima, as componentes
transversais das grandezas físicas se tornarão diferentes. Assim,
por exemplo, a formulação da eletrodinâmica dos táquions assumirá
formas diferentes em cada uma dessas formulações. Somente a experiencia
pode, contudo, decidir qual delas está correta, (isto é, assumindo
que alguma delas esteja). 

Em resumo, os argumentos apresentados acima nos mostram que não é
possível estender as transformações de Lorentz de modo a satisfazer
todos os postulados apresentados na seção \ref{sec: space-time},
a não ser que se introduza quantidades imaginárias, o que não tem
interpretação física. Veremos na seção seguinte que, se não exigirmos
a validade do postulado 1, isto é, que o universo tenha quatro dimensões,
então torna-se possível construir uma teoria dos táquions que satisfaça
os outros postulados.

\section{Uma possível teoria em 6 dimensões\label{sec: Lorentz6}}

Podemos nos perguntar por que encontramos dificuldades em estender
as transformações de Lorentz em um universo de quatro dimensões, se
em duas dimensões essa extensão é quase que direta. Após um pouco
de reflexão, podemos encontrar a resposta. O motivo pelo o qual isso
ocorre reside no fato de que em quatro dimensões temos um número diferente
de dimensões espaciais e temporais. Desde que a experiência nos mostra
que o universo tem três dimensões espaciais (pelo menos), segue que
devemos considerar um universo de (no mínimo) seis dimensões -- três
do tipo tempo, três do tipo espaço%
\footnote{A concepção de um universo em seis dimensões também já foi proposta
antes, conferir \cite{Recami}.%
}. 

\begin{figure}[h]
\begin{center}\includegraphics[scale=0.4]{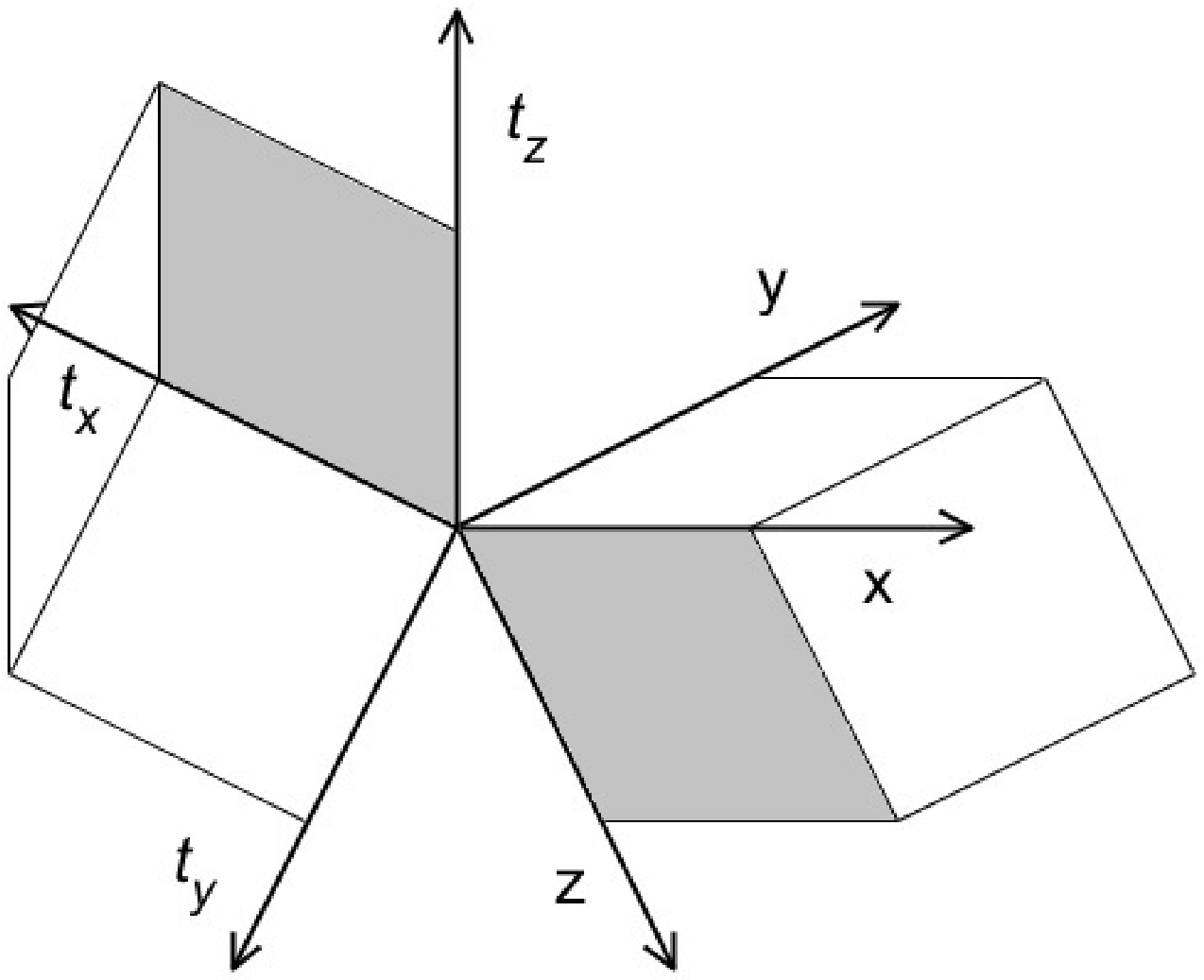}\end{center}

\caption{Ilustração de um universo em 6 dimensões -- ``universos ortogonais.''}

\label{figura: 6}
\end{figure}

Mas o que devemos entender por um universo com três dimensões de tempo?
A interpretação que propomos aqui é a seguinte: embora o universo
tenha três dimensões do tipo tempo, o \emph{tempo físico}, isto é,
o tempo que de fato é medido por um observador, é sempre unidimensional.
Efetivamente, esse tempo físico deve corresponder ao \emph{tempo próprio}
desse observador, ou seja, deve ser determinado pelo comprimento de
sua linha de universo. 

Dessa forma, consideramos também que as outras duas dimensões do tipo
tempo, as que são ortogonais à linha de universo do observador, são
sempre \emph{inacessíveis} a esse observador. Assim, muito embora
o universo seja considerado hexadimensional nessa teoria, o \emph{universo
físico }é sempre quadridimensional. De certa forma, esse universo
a seis dimensões pode ser interpretado como que composto por \emph{dois
universos tridimensionais ortogonais}%
\footnote{O conceito de ``universos paralelos'' já foi, é claro, apresentado
em diversas histórias de ficção científica, entretanto, não me recordo
de nenhuma delas ter introduzido a ideia de ``universos ortogonais.''%
}, conforme está pictoricamente ilustrado na figura \ref{figura: 6}. 

Com essas interpretações, podemos definir a métrica do universo em
seis dimensões pela expressão 
\begin{equation}
ds=\sqrt{\left|c^{2}dt_{x}^{2}+c^{2}dt_{y}^{2}+c^{2}dt_{z}^{2}-dx^{2}-dy^{2}-dz^{2}\right|},
\end{equation}
e a\emph{ métrica física} por 
\begin{equation}
d\sigma=\sqrt{\left|c^{2}dt_{x}^{2}-dx^{2}-dy^{2}-dz^{2}\right|}.\label{eq:MO}
\end{equation}
onde supomos que $t_{x}$ seja o tempo físico medido pelo observador
correspondente (por uma escolha conveniente do sistema de coordenadas
isso, é claro, pode sempre ser feito).

Deixe-nos então mostrar agora como se pode obter as TLE que estejam
de acordo com os postulados apresentados na seção \ref{sec: space-time}.
Para isso, considere um determinado evento de coordenadas $\left(ct_{x},ct_{y},ct_{z},x,y,z\right)$
em relação a um referencial $R$, mas que apenas as coordenadas $\left(ct_{x},x,y,z\right)$
sejam acessíveis a esse referencial. Analogamente, para o referencial
$R'$ deixe $\left(ct_{x}',ct_{y}',ct_{z}',x',y',z'\right)$ serem
as coordenadas daquele mesmo evento, onde agora somente as coordenadas
$\left(ct_{x}',x',y',z'\right)$ são acessíveis a $R'$.

Considere primeiro o caso em que $v<c$. Nesse caso é evidente que
as dimensões acessíveis de $R$ e $R'$ devem ser iguais, já que para
$v=0$ devemos obter a identidade. Assim, obtemos diretamente que
as transformações procuradas consistem nas transformações de Lorentz
usuais, acrescidas das relações 
\begin{equation}
ct_{y}'=ct_{y},\quad ct_{z}'=ct_{z},\quad y'=y,\quad z'=z.
\end{equation}

Considere agora o caso de mais interesse em que $v>c$. Aqui, pelo
contrário, deve haver uma inversão nas dimensões acessíveis de $R$
quando observadas por $R'$. De fato, podemos dizer que os referenciais
$R$ e $R'$ estão, agora, em ``universos ortogonais'' diferentes.
Assim, as coordenadas $ct_{y}$ e $ct_{z}$ devem passar a ser acessíveis
para $R'$, enquanto que as coordenadas $y$ e $z$ devem se tornar
inacessíveis. Note ainda que para a luz se propagar em superfícies
esféricas de velocidade $c$ em ambos os referenciais é necessário
apenas que se tenha $d\sigma=d\sigma'=0$. Contudo, podemos considerar
a hipótese mais forte de que sempre se tenha $ds'=ds=0$ para um raio
de luz. Nesse caso facilmente se verifica que condição anterior é
satisfeita através das relações 
\begin{equation}
ct_{y}'=\pm y,\quad ct_{z}'=\pm z,\quad y'=\pm ct_{y},\quad z'=\pm ct_{z},\label{eq:tytz}
\end{equation}
as quais se somam as expressões 
\begin{equation}
ct_{x}'=\pm\frac{ct_{x}-xv/c}{\sqrt{v^{2}/c^{2}-1}},\quad x'=\pm\frac{x-vt_{x}}{\sqrt{v^{2}/c^{2}-1}},\label{eq:TL4}
\end{equation}
já deduzidas na seção \ref{sec: Lorentz2}, para constituir as TLE
em seis dimensões. Note que agora todas as coordenadas são reais.

Deixe-nos por fim interpretar os resultados obtidos. Em primeiro lugar
é fácil verificar que em ambos os referenciais a luz se propagará
com velocidade $c$. Isso pode ser feito através da substituição direta
de (\ref{eq:tytz}) e (\ref{eq:TL4}) em (\ref{eq:MO}). Além disso,
notemos que nessa formulação, a forma como os táquions são observados
no referencial $R$ pode ser completamente diferente de sua forma
para o referencial $R'$, uma vez que para o referencial $R$ as coordenadas
transversais são dadas por $y=\pm ct_{y}'$ e $z=\pm ct_{z}'$, as
quais não são acessíveis a $R'$. 

Além disso, é interessante notar que se uma fonte de luz for fixada
ao referencial $R'$, então o referencial $R$ verá essa luz se propagando
em superfícies esféricas com velocidade $c$, conforme mostramos,
mas desde que a fonte tem nesse caso uma velocidade maior que $c$,
teremos efetivamente a formação de um cone de Mach, pois a fonte estará
sempre à frente das ondas que emite. Pode-se mostrar então que a superposição
das ondas emitidas por essa fonte superluminal formam duas frentes
de onda com o formato de hiperboloides. A velocidade de grupo dessas
ondas depende da direção, sendo sempre maior que $c$ (exceto na direção
$X$, cuja velocidade é $c$). Isso, no entanto, deve-se puramente
a um fenômeno de interferência, diferentemente do que ocorria nos
casos discutidos na seção anterior. Ondas desse tipo são comumente
chamadas de ``ondas $X$'' e efetivamente são soluções superluminais
das equações de Maxwell \cite{Rodrigues,Recami-3} -- essas ondas,
inclusive, já foram observadas e até mesmo produzidas em laboratório
\cite{bird,Rodrigues,Recami-3,CH1}, o que constitui uma importante
verificação experimental da existência de fenômenos superluminais
na natureza.

\section*{Agradecimentos}

O autor agradece ao Prof. Dr. A. Lima-Santos pela leitura do manuscrito
e sugestões, e também à Fundação de Amparo à Pesquisa do Estado de
São Paulo (FAPESP) pelo suporte financeiro.